\documentclass[11pt,a4paper]{article}

\usepackage{jcappub}
\usepackage{times}
\usepackage{amsfonts}
\usepackage{latexsym}
\usepackage{dcolumn}
\usepackage{bm}
\usepackage{bbm} 
\usepackage{empheq} 
\usepackage{bbold}
\usepackage{epstopdf} 
\newcommand{\ba}{\begin{eqnarray}}
\newcommand{\ea}{\end{eqnarray}}
\newcommand{\be}{\begin{equation}}
\newcommand{\ee}{\end{equation}}
\newcommand{\bea}{\begin{eqnarray}}
\newcommand{\eea}{\end{eqnarray}}
\newcommand{\beq}{\begin{equation}}
\newcommand{\eeq}{\end{equation}}
\newcommand{\beqar}{\begin{eqnarray}}
\newcommand{\eeqar}{\end{eqnarray}}
\newcommand{\beqars}{\begin{eqnarray*}}
\newcommand{\eeqars}{\end{eqnarray*}}
\newcommand{\bc}{\begin{center}}
\newcommand{\ec}{\end{center}}
\newcommand{\ben}{\begin{enumerate}}
\newcommand{\een}{\end{enumerate}}
\newcommand{\bit}{\begin{itemize}}
\newcommand{\eit}{\end{itemize}}
\newcommand{\bw}{\begin{widetext}}
\newcommand{\ew}{\end{widetext}}
\newcommand{\bcl}{\begin{columns}}
\newcommand{\ecl}{\end{columns}}
\newcommand{\ex}{\mbox{e}}
\newcommand{\dd}{\mbox{d}}

\newcommand{\ie}{\emph{i.e.~}}

\newcommand{\eg}{\emph{e.g.~}}
\newcommand{\etc}{\emph{etc.}}

\newcommand{\Hu}{{\cal H}}

\newcommand{\Ka}{{\cal K}}

\newcommand{\GN}{G_{_\mathrm{N}}}

\newcommand{\mrm}[1]{\mathrm{#1}}
\newcommand{\mcl}[1]{\mathcal{#1}}

\newcommand{\ra}{\rangle}
\newcommand{\la}{\langle}
\newcommand{\lb}{\left(}
\newcommand{\rb}{\right)}
\newcommand{\lsb}{\left[}
\newcommand{\rsb}{\right]}
\newcommand{\lcb}{\left\{}
\newcommand{\rcb}{\right\}}
\newcommand{\nn}{\nonumber}

\newcommand{\scpt}{\scriptsize}

\title{Production of non-gaussianities through a positive spatial
curvature bouncing phase}

\author[a]{Xian Gao}

\author[b]{Marc Lilley}

\author[c]{Patrick Peter}

\affiliation[a]{Department of Physics, Tokyo Institute of Technology, 2-12-1 Ookayama, Meguro-ku, Tokyo 152-8551, Japan}

\affiliation[b]{$\mcl{ACGC}$, Department of Mathematics and Applied Mathematics, University of Cape Town, Rondebosch 7701, South Africa}

\affiliation[c]{${\mathcal{G}}{\mathbb{R}}
\varepsilon{\mathbb{C}}{\mathcal{O}}$, Institut d'Astrophysique de Paris, UMR 7095-CNRS, Universit\'{e} Pierre et Marie Curie-Paris 6, 98bis Boulevard Arago, 75014 Paris, France}

\emailAdd{gao(at)th.phys.titech.ac.jp}
\emailAdd{marc.lilley(at)uct.ac.za}
\emailAdd{peter(at)iap.fr}

\date{\today}


\abstract{We compute the level of non-gaussianities produced by a cosmological bouncing phase in the minimal non-singular setup that lies within the context of General Relativity when the matter content consists of a simple scalar field with a standard kinetic term. Such a bouncing phase is obtained by requiring that the spatial sections of the background spacetime be positively curved.  We restrict attention to the close vicinity of the bounce by Taylor expanding the scale factor $a(\eta)$, the scalar field $\phi(\eta)$ and the potential $V(\phi)$ in powers of the conformal time $\eta$ around the bounce.  We find that possibly large non-gaussianities are generically produced at the bounce itself and also discuss which shapes of non-gaussianities are most likely to be produced.}

\begin{document}
\maketitle
\section{Introduction}

Accurate measurements \cite{Ade:2013zuv} of the cosmological parameters within the context of $\Lambda$CDM \cite{Mukhanov:2005sc,PeterUzan2009} have shown that the early epochs of the Universe most probably \cite{Martin:2013nzq} included a phase of primordial inflation \cite{Lemoine:2008zz,Martin:2013tda,Linde:2014nna} (contrary to a recent claim \cite{Ijjas:gauss2013vea} that appears to contradict the Planck team's own conclusions \cite{Ade:2013uln}). The very likely existence of an inflationary period stems from the fact that the primordial spectrum inferred from the data is slightly red, thus excluding the Harrison-Zel'dovich scale-invariant one at the 5-$\sigma$ confidence level, and, more importantly with regards the conclusions of the present work, from the lack of observed non-gaussianities \cite{Ade:2013uln,Ade:2013ydc}.

Non-gaussianities of the observed Cosmic Microwave Background (CMB) and Large-scale Structure (LSS) represent a powerful tool to distinguish among different early universe cosmological models at the origin of structure formation \cite{Bartolo:2004if}. In the past decade, substantial efforts have been made to compute the amplitude and shapes of non-gaussianities present in the statistics of primordial curvature perturbations generated in various inflationary models \cite{Maldacena:2002vr,Chen:2006nt} (see also \eg \cite{Chen:2010xka,Komatsu:2009kd,Komatsu:2010hc} for a review and references therein).  In inflation, although primordial curvature perturbations may be intrinsically non-gaussian, the amplitude in the simplest single field slow-roll inflationary models, for instance, is predicted to be $\ll \mathcal{O}(1)$ \cite{Maldacena:2002vr}.  Because of the nonlinear nature of gravity, the non-gaussianities present in the CMB and LSS also receive contributions from the post-inflationary evolution evaluated to be of order $\mathcal{O}(1)$ \cite{Bartolo:2010qu}.

Non-gaussianities are usually characterized by a set of so-called nonlinear parameters $f_{_\mathrm{NL}}$, which are related to the amplitude of the three-point correlation function of cosmological perturbations.  Assuming the curvature perturbation $\zeta$ to take the form $\zeta = \zeta_\mathrm{lin} + \frac{3}{5}f_{_\mathrm{NL}} \zeta_\mathrm{lin}^2 + \cdots$, where $\zeta_\mathrm{lin}$ is the perturbation at linear order, one finds that the three-point function $\langle \zeta (\bm{x_1}) \zeta (\bm{x_2}) \zeta (\bm{x_3})\rangle$ is proportional to $f_{_\mathrm{NL}}$, the overall amplitude roughly given (at leading order) by the square of the linear power spectrum. The recently
released Planck data have already put stringent constraints on these nonlinear parameters \cite{Ade:2013uln,Ade:2013ydc}.  For instance, at the 1-$\sigma$ level, the so-called local-type non-gaussianity is constrained to be $f_{_\mathrm{NL}}^{\mathrm{(local)}} = 2.7\pm 5.8$, while the equilateral type should not exceed $f_{_\mathrm{NL}}^{\mathrm{(equil)}}= -42\pm 75$ and the orthogonal type $f_{_\mathrm{NL}}^{\mathrm{(ortho)}} = -25\pm 39$.

Most bouncing alternatives to inflation not only suffer from several theoretical drawbacks but also appear to be severely disfavoured by the currently available data \cite{DianaPP2014}\footnote{Bouncing cosmologies would be further disfavoured by the discovery of a large tensor-to-scalar ratio through the measurement of primordial B-mode polarisation in the CMB, as for instance in the BICEP2 experiment~\protect\cite{Ade:2014xna} if confirmed by a refined analysis and by future data.}. Nevertheless, they do represent a possibly viable alternative, and should thus be investigated in full, were it only to enhance our ability to reject their viability by computing explicit predictions and demonstrating their incompatibility with cosmological data.  Thus far, much attention has been paid to generating a scale-invariant curvature spectrum during a contracting phase \cite{Finelli:2001sr,Lehners:2007ac} or through an actual bounce \cite{Pinho:2006ym,Peter:2002cn,Peter:2008qz}, in order to comply with the WMAP and Planck results.  It is worth stressing that a definite prediction can only be made if the spectrum of perturbations produced during a contracting phase is computed after their transfer through the bouncing phase.  Indeed, when this transfer is taken into account, the resulting spectrum is not necessarily equal to the one 
computed in the contracting phase \cite{Martin:2001ue}, and can be modified in a potentially drastic way \cite{Martin:2002ar,Martin:2003bp}\footnote{The spectrum may not always be as dramatically changed, however, in bouncing models realized by non-conventional matter contents, see e.g. \cite{Cai:2008qw} and \cite{Xue:2013bva}.}.

Leaving aside controversies about the actual power spectrum produced in such models, the amount of non-gaussianity that is now constrained to very small values by the recent Planck data \cite{Ade:2013uln,Ade:2013ydc} provides an additional test with the potential to demonstrate the existence of severe discrepancies between observations and theoretical predictions of some classes of cosmological models. This is indeed how classes of inflationary models have been ruled out. Bouncing cosmologies can be similarly scrutinized. Till now, non-gaussianities in bouncing scenarios seem to have attracted much less attention, and in particular, their computation has been, much as has been the case of the power spectrum, largely confined to the contracting phase (see however \cite{Fertig:2013kwa}).  In Ref. \cite{Cai:2009fn} for instance, the production of non-gaussianities has been investigated in the so-called ``matter-bounce'', consisting of a matter-dominated contracting phase followed by a bounce connecting to the currently expanding phase; the choice of such a model stems from the fact that it provides an efficient way of generating a scale-invariant spectrum \cite{Finelli:2001sr}, albeit slightly blue in most cases \cite{Pinho:2006ym,Peter:2006hx}. The authors of \cite{Cai:2009fn} evaluated the non-gaussianity generated by quantum fluctuations during the contracting/collapsing phase, \ie before the occurrence of the bounce.  However, observables are evidently related to perturbations in the subsequent expanding phase. To this end (as has in some cases been calculated in details for the power spectrum) the transfer of non-gaussianities through the bouncing phase has to be computed. 

In the present work, we address this question by providing a detailed calculation of the production of non-gaussianities by the bouncing phase itself.  We attempt to keep the computation as general as possible by making the least possible number of specific assumptions.  We make no further assumptions than those made in the simplest inflationary models.  Indeed, the bouncing phase is obtained within the framework of General Relativity (GR), with a minimally coupled scalar field, with a standard kinetic term, but with positive spatial sections of the background spacetime.  This work is an extension to the study of non-gaussianities of Refs.~\cite{Martin:2003sf,Falciano:2008gt,Lilley:2011ag}.  Just as, for example, post-inflationary contributions to non-gaussianity are due to the intrinsic nonlinearity of gravitation, and can be used to constrain modified gravity models (as, e.g., \cite{Gao:2010um}), the deformation, amplification, suppression or production of non-gaussianities during the bouncing phase are directly related to the details of the bouncing background, and thus can be used to constrain bouncing cosmologies.

As stated above, we work in a minimalistic set-up in which we only require the positivity of the curvature of spatial hypersurfaces. In particular, we make no assumption about the shape of the potential of the scalar field $\phi$.  At linear order in perturbations, many of the relevant observational signatures for such a model have been calculated in a specific setting in which the bounce is followed by an inflationary phase \cite{Cai:2008qb,Lilley:2011ag}\footnote{A spatial-curvature-induced non-singular cosmology that involves a period of inflation post-bounce~\cite{Lilley:2011ag} is one way to satisfy observational constraints on the value of $\Omega_{\mcl K}$, the other being the requirement that the bounce be close to symmetric~\cite{Peter:2008qz}. In a true ``alternative-to-inflation'', one has indeed to assume that the bounce is sufficiently close to being symmetric.  In Section \ref{sec:shapefunctions} onwards we shall restrict the analysis to this latter case, having verified that asymmetric terms will yield sub-dominant contributions to the non-gaussian signal.}.  Here, we make no assumption about the post-bounce cosmological evolution, focus on the bouncing phase itself, and follow Refs~\cite{Martin:2003sf} in expanding the background quantities (scale factor, scalar field, \etc) around the bouncing point up to fourth order in the conformal time $\eta$ in order to obtain an exact expression for second order perturbations at quadratic order in $\eta$. This approach is robust as long as we concentrate on the vicinity of the bouncing point.  We use generic Gaussian statistics as initial conditions 
for first order perturbations on some initial spatial hypersurface at some initial time $\eta_-$ in the contracting phase.  In full generality, there are two contributions to the non-gaussian 
signal in the expanding phase that follows the bounce.  The first is the transfer of existing non-gaussianity generated by quantum fluctuations before the bounce, such as those obtained in \cite{Cai:2009fn}. The other is the non-gaussianity generated during the bouncing phase itself, and which is due to the nonlinearities of gravitation, and controlled by the details of the bounce.  The initial non-gaussianity is determined by the details of interactions in the contracting phase and are thus highly model-dependent.  In this work, we focus on the second contribution, \ie the non-gaussianity generated during the bounce itself, and thus assume a purely gaussian initial state. As we shall see, some sizeable amount of non-gaussianity is generically produced, and its amplitude depends on the shape of the underlying bounce as well as on the timescale of the bounce.

The paper is organised as follows. In Section \ref{sub:background}, we discuss the background cosmology. In Sections \ref{sub:linear} and \ref{sub:initcond} we discuss the evolution of perturbations at the linear level and spend some time discussing our choice of initial conditions at $\eta_-$. In Section \ref{sec:nonlin} we turn to the calculation of the nonlinear evolution through the bounce, and give the explicit expression for the evolution equation of the second order part of the Bardeen potential in a spatially closed Friedmann-Lema\^{\i}tre-Robertson-Walker (FLRW) background. We next write down the general form of the bispectrum (Section \ref{sub:bispectrumgeneral}) and evaluate the production of non-gaussianities in terms of a set of shape functions which naturally appear in the formula for the bispectrum, see Section \ref{sec:shapefunctions}. In the equilateral, squeezed and folded configurations, Sections \ref{sub:equi} to \ref{sub:folded}, we obtain analytic approximations for the entries of a matrix that encodes the production of non-gaussianities across the bounce and that provides an order of magnitude estimate of the parameter $f_{_\mathrm{NL}}$.  Finally, all cumbersome calculational steps are detailed in appendix \ref{EinsteinEqs} while some additional details are provided in appendices \ref{powerspc} to \ref{sourcetermkspace}.

Note that this work is split into two main parts.  The first, from Section \ref{sub:background} to \ref{sub:bispectrumgeneral} included, provides an exact calculation in a general spatially curved FLRW background and also sets up the general framework needed to study the production of non-gaussianities by the bouncing phase.  The second part, from Section \ref{sec:shapefunctions} onwards, restricts the analysis to the case of a symmetric bounce and is in that sense not fully general.  Where appropriate we comment on the inclusion of some level of asymmetry in the bouncing phase (see however footnote 3)
 
\section{Background cosmology and linear perturbations}

\subsection{Background cosmology}
\label{sub:background}

Inflationary cosmology usually starts from an FLRW spacetime with
flat spatial sections. This is justified because during inflation the
spatial curvature term in the Friedmann equations scales as $a^{-2}$
while the inflaton field behaves like a cosmological constant for a
minimum of 50 to 60 $e$-folds. By contrast, in a general bouncing scenario for which the scale factor can become small, the spatial curvature term may not always be negligible, but can dominate in the vicinity of the bouncing point. This is the case when the matter content is made of a single positive energy density fluid: then the sole source of Null Energy Condition (NEC) violation stems from the spatial curvature; this is the case for the model discussed in the present work.

It is important to note that in some models the spatial curvature can be made to be close to vanishing in an ekpyrotic contraction. This is also true for any given initial amount of anisotropy.  In [38], it is shown, in a model with a Bianchi I metric and an ekpyrotic contraction followed by a non-singular bounce obtained using a Galileon-type Lagrangian, that anisotropies can be kept sub-dominant all through the bounce. Although it has not been explicitely demonstrated, one might expect that a similar mechanism could be successful in keeping the spatial curvature sub-dominant all through the bounce.  In such models, the perturbations equations are significantly more complicated than those used in the present work, and include NEC-violating degrees of freedom.  Whether or not these new terms would act as sources of non-gaussianities remains to be calculated.

Given the previous considerations, we  assume the metric to be of the form
\begin{equation}
\dd s^2 = a^2(\eta)\left[-\dd \eta^2 +
\delta_{ij} \dd x^i \dd x^j \left(1+\displaystyle\frac{\Ka}{4}
\delta_{mn} x^m x^n\right)^{-2} \right],
\label{FLRW} 
\end{equation}
where the spatial part depends on the spatial curvature $\Ka$,  With the metric (\ref{FLRW}), 
the Einstein equations read (we set units such that $\hbar=c=8\pi \GN\equiv 1$)
\begin{equation} 
\Hu^2+\Ka =
\frac13\rho a^2 \ \ \ \hbox{and} \ \ \ \ \Hu' =-\frac16a^2 \left(\rho +
3 P\right), 
\label{FriedConf} 
\end{equation} 
where the conformal Hubble rate $\Hu$ is defined by $\Hu\equiv a'/a$, a prime meaning a derivative w.r.t. the conformal time $\eta$ and where the quantities $\rho$ and $P$ stand respectively for the energy density and pressure of the matter content.

Following \cite{Martin:2003sf,Falciano:2008gt,Lilley:2011ag}, we consider the simplest possible model in which a bounce can be accommodated in the framework of GR: $\mcl K=+1$ with the matter content sourcing the Einstein equations (\ref{FriedConf}) taken to be a single scalar field $\phi$ whose dynamics is governed by a canonical kinetic term and a potential $V(\phi)$. The corresponding background equations of motion (\ref{FriedConf}) thus read 
\begin{equation} \frac{\phi'^{2}}{a^{2}}  = 
\frac{2}{a^{2}}\left(\Hu^2-\Hu'+\Ka\right) \ \ \ \ \hbox{and} \ \ \ \ \
-\frac{6}{a^{2}}\mathcal{H}' = -2
V(\phi)\left[1-\frac{\phi'^{2}}{a^{2}V(\phi)}\right],\label{bg_eom}
\end{equation}
the combination of which yields the Klein-Gordon equation $\phi'' +2\mathcal{H} \phi' +a^2 V_{,\phi} =0$. The simplest background cosmology in inflation is obtained from slight deviations from a spatially flat de Sitter cosmology.  Here, instead, the simplest nonsingular background cosmology is obtained from a slight deviation away from de Sitter cosmology in a spacetime of positively curved spatial sections.  The $\mcl K=+1$ de Sitter scale factor reads
\begin{equation}
a(\eta)=a_0\sec\lb\eta\rb\,,
\end{equation}
and two kinds of generic deviations from the de Sitter solution can be introduced. The first is an overall deviation from the de Sitter bouncing timescale through 
\begin{equation}
a(\eta)=a_0\sec\lb\frac{\eta}{\eta_{\mrm c}}\rb,
\label{qdS}
\end{equation}
where $\eta_{\mrm c}=1/(\omega a_0)$, $\omega$ being a dimensionful constant equal to $1/a_0$ in the de Sitter case. It can be checked that the null energy condition is preserved provided $\omega<a_0^{-1}$, so that $\eta_{\mrm c}\geq 1$ \cite{Martin:2003sf}. Such non-de-Sitter bounces thus occur over conformal timescales greater than $\pi$. This first kind of deviation from the de Sitter case is however not sufficient as it yields a constant potential for cosmological perturbations \cite{Martin:2003sf}.

In order to achieve further deviations from de Sitter in the vicinity of the bounce point, we Taylor-expand (\ref{qdS}) and modify the resulting expression by introducing a set of parameters $\lambda_i$. Our working scale-factor then reads
\begin{equation}
a(\eta)=a_0\lsb
1+\frac{1}{2}\lb\frac{\eta}{\eta_{\mrm
c}}\rb^2+\lambda_3\lb\frac{\eta}{\eta_{\mrm c}}\rb^3+\frac{5}{24}\lb
1+\lambda_4\rb\lb\frac{\eta}{\eta_{\mrm c}}\rb^4\rsb.
\label{Taylora}
\end{equation}
The scale factor stays close to de Sitter as long as $\eta_c$ is close to unity, and if the parameters $\lambda_i$ are close to zero.  In this form, this Taylor-expanded metric is fully general, as the constants $\eta_c$ and $\lambda_i$ may take arbitrary values.  In sections (\ref{sec:shapefunctions}) onwards, after having outlined a fully general calculation, we shall restrict our attention to bounces close to the de Sitter bounce.  Such bounces fall into the {\it slow bounce} category in the sense that the scalar field conformal time derivative remains close to zero during the bouncing phase.

Solving the Friedmann equations and the Klein-Gordon equation order byorder in powers of $\eta$, we can write down the constants $a_0$, $\eta_c$ and the $\lambda_i$'s in terms of $\Upsilon\equiv \phi_0'^2/2$ and the successive derivatives of $V(\phi)$ with respect to $\phi$, evaluated at the bounce point. Introducing the following notation, reminiscent of the one used in slow roll inflation,
\begin{equation}
\varepsilon_V=\frac{V'_0}{V_0},\qquad\eta_V=\frac{V_0''}{V_0}\,,
\end{equation}
we find that the coefficients $\lambda_i$ in the Taylor expansion (\ref{Taylora}) can be expressed in terms of $\Upsilon$, $\varepsilon$ and $\eta_V$ through
\be
\eta_c^2=\frac{1}{1-\Upsilon}\,,\nn\qquad a_0=\frac{3-\Upsilon}{V_0}\,,
\ee
and
\be
\lambda_3=\frac{1}{3\sqrt{2}}\frac{(3-\Upsilon)\sqrt{\Upsilon}}{(1-
\Upsilon)^{3/2}}\varepsilon_V\,,\qquad \lambda_4=\frac{4\Upsilon}{5(1-\Upsilon)}-\frac{(3-\Upsilon)^2}{5(1-\Upsilon)^2}\varepsilon_V^2+\frac{2\Upsilon}{5}\frac{3-\Upsilon}{(1-\Upsilon)^2}\eta_V\,.
\ee
The bounce is thus controlled by $\Upsilon$ and the two first derivatives of the potential normalized by $V_0$.  In a similar way, we express the coefficients $\mcl H_0'$, $\mcl H_0''$, $\phi_0''$ and $\phi_0'''$, that appear in the expansions
\begin{equation}
\mcl H(\eta)=\mcl H_0'\eta+\frac{\mcl H_0''}{2}\eta^2\qquad\text{and}\qquad\phi(\eta)=\phi_0+\phi_0'\eta+\frac{\phi''_0}{2}\eta^2+\frac{\phi^{(3)}}{6}\eta^3
\end{equation}
in terms of $\Upsilon$, $\varepsilon_V$ and $\eta_V$. The background cosmology in the neighborhood of the bounce is thus entirely specified by $\Upsilon$, $\varepsilon_V$ and $\eta_V$. Note that in the actual application below (Section \ref{sec:shapefunctions} onwards), we shall concentrate on a symmetric bounce, and thus set $\varepsilon_V$ to zero. Inclusion of some amount of asymmetry during the bounce does not modify our conclusions, except insofar as it induces the production of additional non-gaussianities at next-to-leading order in our computation scheme. What is therefore produced in our symmetric model can thus be understood as the minimum level of non-gaussianities expected in curvature-dominated bounces.

Finally, note that although we have no specific knowledge on the pre-bounce phase, but because the de Sitter solution is an attractor in most theories of gravity, we assume that our bouncing cosmolgy is driven towards the de Sitter attractor.  We will therefore take $\Upsilon$, $\varepsilon_V$ and $\eta_V$, to be small and present the results in Section \ref{sec:shapefunctions} onwards as expansions in these parameters.

\subsection{Linear perturbations}
\label{sub:linear}
We work with the metric in the Poisson gauge (or the generalized Newtonian gauge) and consider only scalar perturbations. The perturbed metric therefore reads
\begin{equation}
\dd s^{2}=g_{\mu\nu}\dd x^{\mu}\dd x^{\nu}=a^{2}\left(-\ex^{2\Phi}\dd\eta^{2}+\ex^{-2\Psi}\gamma_{ij} \dd x^{i}\dd x^{j}\right),
\label{metric_pert_poisson}
\end{equation}
where $\gamma_{ij}$ is the metric on the background spatial hypersurface (\ref{FLRW}). In a homogeneous and isotropic spatial volume, it is convenient to work in spherical coordinates where $\gamma_{ij}$ takes the form
\begin{equation}
\gamma_{ij}\dd x^i\dd x^j = \dd\chi^{2}+\Ka^{-1} \sin^2\left(\sqrt{\Ka}\chi\right)\left(\dd\theta^{2}+\sin^{2}\theta \dd\phi^{2}\right)\,.
\label{eq:spmetric}
\end{equation}
In this work, we expand all quantities up to the second order as
\begin{equation}
X\left(\bm{x},\eta\right)=X_{(1)}\left(\bm{x},\eta\right)+\frac{1}{2}X_{(2)}\left(\bm{x},\eta\right)+\cdots,
\label{eq:fieldexp}
\end{equation}
where $X$ stands for $\Phi$, $\Psi$ and $\delta\phi$, \etc, and $X_{(1)}$ and $X_{(2)}$ are the first and second order perturbations respectively. In a universe filled with single scalar field matter, the equation of motion for the linear perturbation $\Psi_{(1)}$ is well-known and reads (see Appendix \ref{EinsteinEqs})
\begin{equation}
\Psi_{(1)}''+F\left(\eta\right)\Psi_{(1)}'-\bar{\nabla}^{2}\Psi_{(1)}+W\left(\eta\right)\Psi_{(1)}=0,
\label{Phi_1_eom}
\end{equation}
where $\bar\nabla^2$ is the Laplacian w.r.t. the spatial metric (\ref{eq:spmetric}) and where
\begin{equation} F(\eta) =
2\left(\mathcal{H}-\frac{\bar{\phi}''}{\bar{\phi}'}\right),\qquad W(\eta) = 2\left(\mathcal{H}'-\mathcal{H}\frac{\bar{\phi}''}{\bar{\phi}'}-2\Ka \right).
\end{equation}
Since $\mcl K=+1$, and in order to compute the spectral matrix $\bm P$ and the bispectrum $\mcl B_{\Psi}$ which we introduce in the following sections, the first order perturbation is decomposed on the three-sphere in terms of the set of hyperspherical harmonics $Q_{\ell m n}(\chi,\theta,\varphi)$ as follows:
\begin{equation}
\Psi_{(1)}(\bm x,\eta)=\sum_{\ell mn}\Psi_{\ell m n}(\eta)Q_{\ell mn}(\chi,\theta,\varphi)
\quad\text{with}\quad Q_{\ell mn}(\chi,\theta,\varphi)=R_{\ell n}(\chi)Y_{\ell m}(\theta,\varphi),
\end{equation}
with $R_{\ell n}(\chi)$ defined in Appendix \ref{FormFactor}.

\subsection{Stochastic initial conditions}
\label{sub:initcond}
In order to specify the set of initial conditions for the first order 
perturbation $\Psi_{(1)}$ prior to the bounce, let us consider some 
spatial hypersurface $\mcl M$ at some initial time $\eta_-$ before the bounce. 
On $\mcl M$, let us assume that the initial conditions can be written as two
\emph{classical} gaussian random fields, one for $\Psi$ and one for its
time derivative at $\eta_-$: 
\begin{equation} \left[\begin{array}{c}
\Psi_{(1)}\left(\bm{k},\eta_{-}\right)\\
\Psi_{(1)}'\left(\bm{k},\eta_{-}\right) \end{array}\right] \equiv \left[
\begin{array}{c} \hat{x}_{1}\left(\bm{k}\right)\\
\hat{x}_{2}\left(\bm{k}\right) \end{array} \right].
\label{eq:Psi1vec}
\end{equation}
In this work, we shall treat the set $\{\hat{x}_i\}$ as
the initial conditions and shall make use of their two-point
correlations which we define as (see Appendix \ref{powerspc})
\begin{equation}
\left\langle
\hat{x}_{i}\left(\bm{k}\right)\hat{x}_{j}\left(\bm{k}'\right)
\right\rangle \equiv\delta_{\bm{k},\bm{k}'} P_{ij}\left(k\right)
\qquad\text{with}\qquad \delta_{\bm{k},\bm{k}'} \equiv
\delta_{nn'}\delta_{\ell \ell'} \delta_{mm'} 
\label{P_ij_def}
\end{equation}
where $P_{12}\neq 0$ in the general case.  The matrix $\bm{P}$ is the
generalization of the usual power spectrum, and is determined by the
cosmological history that precedes the bouncing phase. The power
spectrum for $\Psi_{(1)}$ is of course simply $P_{11}$.

In order to compute the evolution of the second order perturbations
through the bouncing phase, we use the solution of the dynamical
equation governing the evolution of $\Psi_{(1)}(\bm k,\eta)$,
namely
\be
\Psi_{(1)}(\bm k,\eta)=\hat{a}_{1}\left(\bm{k}\right)\psi_{1}\left(k,\eta\right)+\hat{a}_2\left(\bm{k}\right)\psi_{2}\left(k,\eta\right).
\label{Psi_ran_dec}
\ee
The first order perturbation $\Psi_{(1)}$ is a real function of time and space and thus satisfies the reality condition $\Psi_{(1)}^{\ast}\left(\bm{k},\eta\right)=\Psi_{(1)}\left(-\bm{k},\eta\right)$. The mode functions $\psi_1(k,\eta)$ and $\psi_2(k,\eta)$ are two linearly-independent solutions of (\ref{Phi_1_eom}), while $\hat{a}_1$ and $\hat{a}_2$ are two complex random variables
which encode the stochastic initial conditions.  The $\hat{a}_i$'s can be written in terms of the $\hat{x}_i$'s  as
\begin{equation}
\hat{a}_i = F_{ij} \hat{x}_j, \qquad \text{with}\qquad \bm{F}\equiv
\left[\begin{array}{cc} \psi_{1}(k,\eta_{-}) & \psi_2(k,\eta_{-})\\
\psi_{1}'(k,\eta_{-}) & \psi_{2}'(k,\eta_{-}) \end{array}\right]^{-1}.
\label{a_x_rel}
\end{equation}
If $\hat{x}_1$ and $\hat{x}_2$ are independent normal random fields, then
$\hat{a}_1$ and $\hat{a}_2$ are said to be jointly normal. The statistics of 
$\hat{a}_1$ and $\hat{a}_2$ are encoded in their two
point-functions
\begin{equation}
\left\langle
\hat{a}_{i}\left(\bm{k}\right)\hat{a}_{j}\left(\bm{k}'\right)
\right\rangle \equiv \delta_{\bm{k},\bm{k}'}
A_{ij}\left(k\right)\quad\text{with}\quad \delta_{\bm{k},\bm{k}'} \equiv
\delta_{nn'}\delta_{ll'} \delta_{mm'},
\label{a_2pf}
\end{equation}
where the matrix elements $A_{ij}$ are determined by the correlation functions of the random variables $\hat{x}_i$'s, the matrix $\bm A$ being given by
\begin{equation}
\bm{A} =\bm{F} \bm{P} \bm{F}^{T}.
\end{equation}
Note that due to the reality condition on $\Psi_{(1)}$, not all components of $A_{ij}$ defined in (\ref{a_2pf})
are independent.  Note also that the initial conditions on the spatial hypersurface $\mcl M$ at time $\eta_-$ are determined by the cosmological history for times $\eta<\eta_-$. We make no assumption about the cosmology prior to the bounce so that the matrix $\bm A$ or equivalently the matrix $\bm P$ are specified no further.

During and after the bounce, the power spectrum of
$\Psi_{(1)}$ is given in terms of the mode vector $\bm{\psi} = \left(
\begin{array}{c} \psi_1 \cr \psi_2 \end{array} \right)$ and the $2\times
2$ matrix $\bm{A}$ as
\begin{equation}
P_{\Psi} (k,\eta) = \bm\psi^{T}\left(k,\eta\right) \bm{A}(k)\bm\psi\left(k,\eta\right)
\equiv \bm v^{T}\left(k,\eta\right)  \bm{P}(k)  \bm v\left(k,\eta\right),
\label{Psi_spectrum}
\end{equation}
where $\bm{P}$ was defined in (\ref{P_ij_def}) and where
\begin{equation}
\bm v\left(k,\eta\right)\equiv
\bm{F}^{T}(k)\bm\psi\left(k,\eta\right)\,.
\label{v_def}
\end{equation}
More explicitly, we have $\bm v=\lb \begin{array}{c} v_1 \cr v_2\end{array}\rb$ with
\begin{equation}
\left[\begin{array}{c}
v_{1}\left(k,\eta\right)\\ v_{2}\left(k,\eta\right)
\end{array}\right]=\frac{1}{\mathcal{W}_{\psi}\left(k,\eta_{-}\right)}
\left[\begin{array}{c}
\psi_{1}\left(k,\eta\right)\psi_{2}'\left(k,\eta_{-}\right)-\psi_{1}'
\left(k,\eta_{-}\right)\psi_{2}\left(k,\eta\right)\\
-\psi_{1}\left(k,\eta\right)\psi_{2}\left(k,\eta_{-}\right)+\psi_{1}
\left(k,\eta_{-}\right)\psi_{2}\left(k,\eta\right) \end{array}\right],
\label{v_expl}
\end{equation}
where $\mcl W_{\psi}(k,\eta)$ is the Wronskian of $\psi_{1,2}$ and reads
\begin{equation}
\mcl W_{\psi}(k,\eta) \equiv
\psi_{1}\left(k,\eta\right)\psi_{2}'\left(k,\eta\right)-\psi_{1}'\left(k
,\eta\right)\psi_{2}\left(k,\eta\right).
\end{equation}
Note that neither $\{v_1,v_2\}$ nor $\bm P_{\Psi}(k,\eta)$ depend on the choice of solutions $\{\psi_1,\psi_2\}$.  Furthermore, $\{v_1,v_2\}$ is the set of solutions of (\ref{Phi_1_eom}) with initial
conditions at $\eta_{-}$ given by
\begin{equation}
v(k,\eta_-) = \left(\begin{array}{c}1\\ 0\end{array}\right)\qquad \hbox{and} \ \ \ \ \
v'(k,\eta_-) = \left(\begin{array}{c}0\\ 1\end{array}\right)\,.
\label{v_ini_cond}
\end{equation}
This implies that with this choice of mode functions, the initial conditions on the initial hypersurface $\mcl M$ are entirely encoded in the spectral matrix $\bm P$.

In this work, we focus on the behaviour of perturbations around the
bouncing point and thus solve for $v_1$ and $v_2$ up to quadratic order in the neighborhood of $\eta=0$. We find that 
\begin{equation}
v_{i}(\eta) = c_{i,0} +c_{i,1} \eta +
\frac{1}{2} c_{i,2} \eta^2,\qquad i=1,2, 
\label{v_sol}
\end{equation}
where the coefficients $c_{i,0}$, $c_{i,1}$ and $c_{i,2}$ are functions of $k$.
Plugging (\ref{v_sol}) into (\ref{Phi_1_eom}) and setting $\eta=0$ yields
\begin{equation}
c_{i,2}+F_{0}\, c_{i,1}+\left(k^{2} + W_{0}\right) c_{i,0}=0,
\end{equation}
with
\bea
F_{0}&\equiv&
F\left(\eta=0\right)=-2\frac{\bar{\phi}_{0}''}{\bar{\phi}_{0}'}=(
\Upsilon-3)\sqrt{\frac{2}{\Upsilon}}\varepsilon_V\,,\\ 
W_{0}&\equiv&
W\left(\eta=0\right)=2\left(\mathcal{H}_{0}'-2\Ka\right)=2-2\Upsilon-4
\Ka,
\eea
where we have used the fact that $\mathcal{H}(\eta=0)=
0$, and where the subscript ``0'' denotes quantities evaluated at the bounce point.
Together with the initial conditions (\ref{v_ini_cond}),
we immediately get
\begin{eqnarray} 
c_{1,0} & = &
\frac{2-2F_{0}\eta_{-}}{-2F_{0}\eta_{-}+\eta_{-}^{2}k^{2}+\eta_{-}^{2}
W_{0}+2},\label{c10}\\ c_{1,1} & = &
\frac{2\eta_{-}\left(k^{2}+W_{0}\right)}{-2F_{0}\eta_{-}+\eta_{-}^{2}k^
{2}+\eta_{-}^{2}W_{0}+2},\label{c11}\\ 
c_{1,2} & = &
-\frac{2\left(k^{2}+W_{0}\right)}{-2F_{0}\eta_{-}+\eta_{-}^{2}k^{2}+
\eta_{-}^{2}W_{0}+2},\label{c12}
\end{eqnarray}
and
\begin{eqnarray}
c_{2,0} & = &
\frac{\eta_{-}\left(F_{0}\eta_{-}-2\right)}{-2F_{0}\eta_{-}+\eta_{-}^{2
}k^{2}+\eta_{-}^{2}W_{0}+2},\label{c20}\\ 
c_{2,1} & = &
-\frac{\eta_{-}^{2}k^{2}+\eta_{-}^{2}W_{0}-2}{-2F_{0}\eta_{-}+\eta_{-}^
{2}k^{2}+\eta_{-}^{2}W_{0}+2},\label{c21}\\ 
c_{2,2} & = &
\frac{2\left(\eta_{-}\left(k^{2}+W_{0}\right)-F_{0}\right)}{-2F_{0}
\eta_{-}+\eta_{-}^{2}k^{2}+\eta_{-}^{2}W_{0}+2}.
\label{c22}
\end{eqnarray}
In Section \ref{sec:bispectrum}, we shall use these expressions to compute the shapes and amplitudes of non-gaussianities induced by the bouncing phase.

\section{Nonlinear evolution of perturbations through the bounce}
\label{sec:nonlin}
The equation of motion for $\Psi_{(2)}$ in real space reads
\begin{equation}
\Psi_{(2)}''+2\left(\mathcal{H}-\frac{\bar{\phi}''}{\bar{\phi}'}\right)
\Psi_{(2)}'-\bar{\nabla}^{2}\Psi_{(2)}+2\left(\mathcal{H}'-2\Ka-
\mathcal{H}\frac{\bar{\phi}''}{\bar{\phi}'}\right)\Psi_{(2)}=\mathcal{S}
_{(2)}, 
\label{Psi_2_eom}
\end{equation}
where the source term $\mathcal{S}_{(2)}$ is
given by
\begin{eqnarray} \mathcal{S}_{(2)} & = &
4\left(2\mathcal{H}^{2}-\mathcal{H}'+2\mathcal{H}\frac{\bar{\phi}''}{
\bar{\phi}'}+6\Ka\right)\Psi_{(1)}^{2}+8\Psi_{(1)}'^{2}+8\left(2\mathcal
{H}+\frac{\bar{\phi}''}{\bar{\phi}'}\right)\Psi_{(1)}\Psi_{(1)}'+8\Psi_
{(1)}\bar{\nabla}^{2}\Psi_{(1)}-\frac{4}{3}\left(\bar{\nabla}_{i}\Psi_{
(1)}\right)^{2}\nonumber\\ &  &
-\left[2\left(2\mathcal{H}^{2}-\mathcal{H}'\right)-\frac{\bar{\phi}'''}
{\bar{\phi}'}\right]\phi_{(1)}^{2}-\frac{2}{3}\left(\bar{\nabla}_{i}
\phi_{(1)}\right)^{2}-2\left(\frac{\bar{\phi}''}{\bar{\phi}'}+2\mathcal{
H}\right)\bar{\nabla}^{-2}\bar{\nabla}^{i}\left(2\Psi_{(1)}'\bar{\nabla
}_{i}\Psi_{(1)}+\phi_{(1)}'\bar{\nabla}_{i}\phi_{(1)}\right)\nonumber\\
&  &
+\left[2\left(\mathcal{H}'-\mathcal{H}\frac{\bar{\phi}''}{\bar{\phi}'}
\right)+\frac{1}{3}\bar{\nabla}^{2}\right]\left[2F\left(\Psi_{(1)}\right
)+F\left(\phi_{(1)}\right)\right]+\mathcal{H}\left[2F\left(\Psi_{(1)}
\right)+F\left(\phi_{(1)}\right)\right]',\label{source_2nd}
\end{eqnarray}
with
\begin{equation}
F\left(X\right)=\left(\bar{\nabla}^{2}\bar{\nabla}^{2}+3\Ka\bar{\nabla}
^{2}\right)^{-1}\bar{\nabla}_{i}\bar{\nabla}^{j}\left[3\bar{
\nabla}^{i}X\bar{\nabla}_{j}X-\delta_{j}^{i}\left(\bar{\nabla}_{k}X
\right)^{2}\right].
\end{equation}
The reader is referred to Appendix \ref{EinsteinEqs} for the details of the calculation. In
$k$-space, $\mathcal{S}_{(2)}$ can be written in the compact form
\begin{equation}
\mathcal{S}_{(2)}\left(\bm{k},\eta\right)=\sum_{\bm{p}_{1},\bm{p}_{2}}
\mathcal{G}_{\bm{k},\bm{p}_{1},\bm{p}_{2}}\,\tilde{\Sigma}_{ij}\left(k,
p_{1},p_{2};\eta\right)\hat{a}_{i}\left(\bm{p}_{1}\right)\hat{a}_{j}
\left(\bm{p}_{2}\right),
\label{source_a1a2_comp}
\end{equation}
where $\mathcal{G}_{\bm{k},\bm{p_1},\bm{p_2}}$ is a geometrical form factor
defined in Appendix \ref{FormFactor} (recall that we are working in an FLRW background with positvely
curved spatial hypersurfaces, so that the usual flat-space Fourier integrals 
are replaced by discrete sums over hyperspherical harmonics, see Appendix \ref{powerspc}), and where
\begin{eqnarray}
\tilde{\Sigma}_{ij}\left(k,p_{1},p_{2};\eta\right)&=&\mathcal{C}_{1}
\left(k,p_{1},p_{2};\eta\right) \psi_{i}\left(p_{1},\eta\right)
\psi_{j}\left(p_{2},\eta\right)+\nn\\
&&\mathcal{C}_{2}\left(k,p_{1},p_{2};\eta\right)\psi_{i}\left(p_{1},
\eta\right)\psi_{j}'\left(p_{2},\eta\right)+\mathcal{C}_{3}\left(k,p_{1}
,p_{2};\eta\right)\psi_{i}'\left(p_{1},\eta\right)\psi_{j}'\left(p_{2},
\eta\right).
\label{Sigma_def}
\end{eqnarray}
with the coefficients $\mcl C_1$, $\mcl C_2$ and $\mcl C_3$ as given in Appendix \ref{sourcetermkspace}.
As already pointed out, it is more convenient to
rewrite the source term in terms of the Gaussian variables
$\hat{x}_i(\bm{k})$ and mode functions $v(k,\eta)$ defined in
(\ref{v_def}). This yields 
\begin{equation}
\mathcal{S}_{(2)}\left(\bm{k},\eta\right)=\sum_{\bm{p}_{1},\bm{p}_{2}}
\mathcal{G}_{\bm{k},\bm{p}_{1},\bm{p}_{2}}\,\Sigma_{ij}\left(k,p_{1},p_{
2};\eta\right)\hat{x}_{i}\left(\bm{p}_{1}\right)\hat{x}_{j}\left(\bm{p}
_{2}\right), 
\end{equation} 
where 
\begin{eqnarray}
\Sigma_{ij}\left(k,p_{1},p_{2};\eta\right) & = &
F_{ki}\tilde{\Sigma}_{kl}F_{lj}\nonumber \\ & = &
\mathcal{C}_{1}\left(k,p_{1},p_{2};\eta\right)v_{i}\left(p_{1},\eta
\right)v_{j}\left(p_{2},\eta\right)+\nonumber \\ &  &
\mathcal{C}_{2}\left(k,p_{1},p_{2};\eta\right)v_{i}\left(p_{1},\eta
\right)v_{j}'\left(p_{2},\eta\right)+\mathcal{C}_{3}\left(k,p_{1},p_{2};
\eta\right)v_{i}'\left(p_{1},\eta\right)v_{j}'\left(p_{2},\eta\right).
\label{Sigma_red_def}
\end{eqnarray} 
From (\ref{Psi_2_eom}), the general solution for $\Psi_{(2)}$ is
\begin{equation}
\Psi_{(2)}\left(\bm{k},\eta\right)=
\Psi_{(2)}^{(0)}\left(\bm{k},\eta\right)
+\sum_{\bm{p}_1,\bm{p}_2}\mathcal{G}_{\bm{k},\bm{p}_{1},\bm{p}_{2}}\,
\Pi_{ij}\left(k,p_{1},p_{2};\eta\right)
\hat{x}_{i}\left(\bm{p}_{1}\right)\hat{x}_{j}\left(\bm{p}_{2}\right),
\label{eq:Psi2general}
\end{equation}
where
$\Psi_{(2)}^{(0)}\left(\bm{k},\eta\right)$ is a particular solution for
$\Psi_{(2)}$ without the source term and where
\begin{equation}
\Pi_{ij}\left(k,p_{1},p_{2};\eta\right)\equiv\int_{\eta_-}^{\eta}\dd
\eta'G\left(k,\eta,\eta'\right)\Sigma_{ij}\left(k,p_{1},p_{2};\eta'
\right),
\label{Pi_def}
\end{equation}
the Green's function being given by
\begin{equation}
G(k,\eta,\eta')\equiv\frac{-v_{1}\left(k,\eta\right)v_{2}\left(k,\eta'
\right)+v_{2}\left(k,\eta\right)v_{1}\left(k,\eta'\right)} {\mcl W_v(k,\eta')}\,,
\end{equation}
and the Wronskian $\mcl W_v(k,\eta')$ by
\begin{equation}
\mcl W_v(k,\eta')=v_{1}\left(k,\eta'\right)v_{2}'\left(k,\eta'\right)-v_{1}'
\left(k,\eta'\right)v_{2}\left(k,\eta'\right).
\end{equation}
Up to now, the calculation is rather general, and can be applied to the non-linear
evolution of $\Psi$ up to second order for an arbitrary cosmological evolution in a spatially closed FLRW background\footnote{Note that the same techniques have been applied earlier to compute the non-linear transfer of the gravitational potential into the temperature anisotropies on large scales \cite{Bartolo:2010qu,Bartolo:2004ty,DAmico:2007iw,Boubekeur:2009uk,Gao:2010ti}.  It was found that the non-linearities of gravitation would
only contribute $\mathcal{O}(1)$ non-gaussianities during the standard big-bang
evolution, although these can be amplified in a modified theory of gravity
\cite{Gao:2010um}. Our work demonstrates that bouncing cosmologies provide
another possibility to enhance nonlinearities.}.  We now evaluate
the elements of the matrix $\bm \Pi$ which is defined in (\ref{Pi_def}) and
encode the influence of \emph{non-linearities} on the
statistics of $\Psi$.  This requires a concrete cosmology both at the level of the background and at the level of first order perturbations which we take to be the ones introduced in Sections \ref{sub:background} to \ref{sub:initcond}.

\section{Bispectrum formula, non-gaussian shapes and amplitudes}
\label{sec:bispectrum}
\subsection{The general form of the bispectrum and parameter $f_{_\mrm{NL}}$}
\label{sub:bispectrumgeneral}
Let us define the bispectrum $\mcl B_{\Psi}(k_1,k_2,k_3;\eta)$ in terms of the three-point function
through the relation
\begin{equation}
\left\langle
\Psi\left(\bm{k}_{1},\eta\right)\Psi\left(\bm{k}_{2},\eta\right)\Psi
\left(\bm{k}_{3},\eta\right)\right\rangle \equiv\frac{1}{2}\mcl
G_{\bm{k}_{1}\bm{k}_{2}\bm{k}_{3}}\mcl
B_{\Psi}\left(k_{1},k_{2},k_{3};\eta\right)
\end{equation}
where $\mcl G_{\bm{k}_{1}\bm{k}_{2}\bm{k}_{3}}$ is the geometric form factor
introduced before and where we have included a factor $1/2$ for
convenience.  Given that the perturbation $\Psi$ is expanded according to
(\ref{eq:fieldexp}), the leading (second order) contribution to the
three-point function is given by
\begin{equation}
\left\langle
\Psi\left(\bm{k}_{1},\eta\right)\Psi\left(\bm{k}_{2},\eta\right)\Psi
\left(\bm{k}_{3},\eta\right)\right\rangle=\frac{1}{2}\left\langle
\Psi_{(1)}\left(\bm{k}_{1},\eta\right)\Psi_{(1)}\left(\bm{k}_{2},\eta
\right)\Psi_{(2)}\left(\bm{k}_{3},\eta\right)\right\rangle+\text{2
perms}.
\end{equation}
Let us focus our attention on the {\it production} of
second order perturbations during the bouncing phase. To that end we neglect $\Psi_{(2)}^{(0)}$ and write
(\ref{eq:Psi1vec}) and (\ref{eq:Psi2general}) in vector form as
\begin{equation}
\Psi_{(1)}(\bm k,\eta)=\bm v(k)^T\hat{\bm x}(\bm k)=\hat{\bm{x}}(\bm
k)^T\bm v(k),\qquad \Psi_{(2)}\left(\bm{k},\eta\right)=
\sum_{\bm{p}_1,\bm{p}_2}\mathcal{G}_
{\bm{k_3},\bm{p}_{1},\bm{p}_{2}}\,\hat{\bm x}(\bm p_1)^T\bm
\Pi(k_3,p_1,p_2)\hat{\bm x}(\bm p_2).
\end{equation}
As made clear in Section \ref{sub:initcond}, the statistics of $\Psi_{(1)}$ are assumed Gaussian.  The three-point function then becomes
\bea
\left\langle
\Psi\left(\bm{k}_{1},\eta\right)\Psi\left(\bm{k}_{2},\eta\right)\Psi
\left(\bm{k}_{3},\eta\right)\right\rangle&=&\frac{1}{2}\la \bm
v^T(k_1)\hat{\bm x}(\bm k_1)
\sum_{\bm{p}_1,\bm{p}_2}\mathcal{G}_{\bm{k_3},\bm{p}_{1},\bm{p}_{2}}\,
\hat{\bm x}(\bm p_1)^T\bm \Pi(k_3,p_1,p_2)\hat{\bm x}(\bm p_2)  
\hat{\bm{x}}(\bm k_2)^T\bm v(k_2)\ra+\text{2 perms}\nn\\ &=&\frac{1}{2}
\sum_{\bm{p}_1,\bm{p}_2}\mathcal{G}_{\bm{k_3},\bm{p}_{1},\bm{p}_{2}} \bm
v^T(k_1)\,\la\hat{\bm x}(\bm k_1)\hat{\bm x}(\bm p_1)^T
\bm\Pi(k_3,p_1,p_2)\hat{\bm x}(\bm p_2)\hat{\bm{x}}(\bm k_2)^T\ra\bm
v(k_2)+\text{2 perms}\nn\\ &=&\frac{1}{2}\mcl G_{k_3,k_1,k_2} \bm
v^T(k_1)\bm P(k_1)\bm{\Pi}(k_3,k_1,k_2)\bm P(k_2)\bm v(k_2)+\text{5
perms},
\eea
where we have used Wick theorem and the definition of the
spectral matrix $\bm P$ defined in (\ref{P_ij_def}) in going from the second to the
third line. The leading order contributions to the bispectrum thus take
the form
\begin{equation}
B_{\Psi}\left(k_{1},k_{2},k_{3};\eta\right)=v^{T}\left(k_{1},\eta\right)
\bm{P}\left(k_{1}\right){\bm{\Pi}}\left(k_{3},k_{1},k_{2};\eta\right)\bm
{P}\left(k_{2}\right)v\left(k_{2},\eta\right)+5\,\mathrm{perms},
\end{equation} 
where ${\bm{\Pi}}$ is defined in (\ref{Pi_def}).  The
amplitude of the bispectrum is popularly characterized by the so-called
dimensionless nonlinear parameter $f_{\mathrm{NL}}$, defined by
\cite{Komatsu:2001rj}
\begin{eqnarray}
f_{_\mathrm{NL}} & \equiv &
\frac{5}{6}\frac{\mcl
B_{\Psi}(k_{1},k_{2},k_{3};\eta)}{P_{\Psi}(k_{1})P_{\Psi}(k_{2})+P_{\Psi
}(k_{2})P_{\Psi}(k_{3})+P_{\Psi}(k_{3})P_{\Psi}(k_{1})}\nonumber \\ 
& = &
\frac{5}{6}\frac{v^{T}\left(k_{1},\eta\right)\bm{P}\left(k_{1}\right)\bm
{\Pi}\left(k_{3},k_{1},k_{2};\eta\right)\bm{P}\left(k_{2}\right)v\left(
k_{2},\eta\right)+5\,\mathrm{perms}}{v^{T}\left(k_{1},\eta\right)\bm{P}
\left(k_{1}\right)v\left(k_{1},\eta\right)v^{T}\left(k_{2},\eta\right)\bm
{P}\left(k_{2}\right)v\left(k_{2},\eta\right)+(2\rightarrow
3)+(1\rightarrow 3)}.
\label{fNL_def}
\end{eqnarray}
This is the main {\it exact} result of this work.  Before proceeding, it is worth remembering that the spectral matrix $\bm{P}$ is not known. Therefore, only the ``sourcing'' factor of non-gaussianities across the bounce can be computed, that is, the matrix elements of $\bm\Pi$.

Although there are a total of six terms
in each matrix element of $\bm\Pi$, it is useful to note that the
off-diagonal elements of the matrix $\bm \Pi$ are symmetric under the
following simultaneous index permutations
\begin{equation}
\Pi_{12}(k_i,k_j,k_l)=\Pi_{21}(k_i,k_l,k_j).
\label{eq:symm}
\end{equation}
This reduces the number of distinct terms in $\bm\Pi$ to 18.  In the equilateral
configuration in which all three wavenumbers are equal, there is just
one contribution to each of the three distinct matrix elements in
$\bm\Pi$.  In both the folded ($k_i=k_j=k$ and $k_{\ell}=2k$) and squeezed ($k_i=k_j=k$ and $k_{\ell}=p\ll k$) configurations for which two
of the three wavenumbers are equal, there are just three distinct terms in each matrix element of $\bm\Pi$.  Furthermore, the following additional symmetry applies
\bea
\Pi_{ij}(k,p,k)-\Pi_{ij}(k,k,p)=0
\quad\text{if $i=j$}
\eea
while it remains that
\bea
&&\left.
\begin{array}{c} \Pi_{ij}(k,p,k)-\Pi_{ij}(k,k,p)\ne 0 \end{array}
\rcb\quad\text{if $i\ne j$}\\
&&\left. \begin{array}{c}
\Pi_{ij}(p,k,k)-\Pi_{ij}(k,p,k) \ne 0\\ \Pi_{ij}(p,k,k)-\Pi_{ij}(k,k,p)
\ne 0 \end{array} \rcb \quad\text{for any $i$ and $j$}\\
&&\left.
\begin{array}{c} \Pi_{ij}(k,p,k)-\Pi_{ji}(k,p,k)\ne0\\
\Pi_{ij}(p,k,k)-\Pi_{ji}(k,p,k)\ne0\\
\end{array} \rcb \quad\text{for any $i \ne j$}.
\eea
These considerations leave a total of just seven distinct contributions to the matrix elements of $\bm \Pi$.

In the following four sections, we compute the contributions to the matrix elements of $\bm \Pi$ in various cases. Given Eq. (\ref{fNL_def}), we expect those matrix elements to provide an order of magnitude estimate of the parameter $f_{_\mathrm{NL}}$. These results hold for {\it symmetric bounces only} -- but as already mentioned in the introduction, and made a little bit more precise below at the end of Section \ref{sec:shapefunctions}, the asymmetry contributes subleading effects in the immediate vicinity of the bounce, and only leads to an enhanced non-gaussian signal.  The following calculations are performed in an expansion in the parameters $\Upsilon$ and $\eta_V$. Assuming these parameters to be small implies that the bounce is close to a de Sitter bounce and that we are assuming the potential to be a smooth and slowly varying function of $\phi$.  Since we focus on the effect of the bouncing phase, it is most natural to evaluate the elements of the amplification matrix from an initial equal-time spatial hypersurface at $\eta_-=-\eta_c=-\sqrt{1/(1-\Upsilon)}$ to a final equal-time spatial hypersurface at  $\eta_+=\eta_c=\sqrt{1/(1-\Upsilon)}$.

In Section \ref{sec:shapefunctions}, we first express the matrix elements of $\bm \Pi$ in terms of a set of shape functions which suggest that the non-gaussianities produced by the bouncing phase are peaked in (bi-)linear combinations of the equilateral and squeezed shapes. We then repeat the calculation specifically in the equilateral, the squeezed, and folded shapes in Sections \ref{sub:equi} to \ref{sub:folded} by further assuming that $k\gg 1 $.  This is justified by the fact that the wavenumbers of observable interest today are within the interval $10^2\lesssim k\lesssim 10^8$ for $\Omega_{\mcl K}\leq 10^{-2}$.  

\subsection{Bispectrum contributions in terms of shape functions}
\label{sec:shapefunctions}
The shape of the bispectrum, \ie its dependence on momentum
configurations, is a function of the details of the physical theory at the origin of the non-gaussian signal.  In a non-singular bouncing cosmology obtained using a single minimally coupled scalar field with a standard kinetic term and a smooth potential, but with an unknown initial perturbation spectrum, the curvature of spatial sections of the background spacetime, the arbitrary (\ie non-standard) initial state and the violation of slow roll in the vicinity of the bounce are distinct features from which non-linearities may become large and the production of non-gaussianities might be expected.  Before proceeding to the quantitative characterization of non-gaussianities across the bouncing phase, let us recall the shapes and amplitudes of the primordial bispectrum generated in inflationary scenarios. As is well-known, standard kinetic term single-field inflationary models predict a negligible (local type) bispectrum in the squeezed limit \cite{Maldacena:2002vr,Creminelli:2004yq}. This is consistent with the latest observational constraints \cite{Ade:2013ydc}, thus explaining why such models are currently largely favoured. Large local type non-gaussianity can be produced in models with either non-Bunch-Davies (BD) vacuum initial conditions (see \eg \cite{Ashoorioon:2010xg,Creminelli:2011rh,Agarwal:2012mq} and references
therein) or containing an early non-attractor phase \cite{Namjoo:2012aa,Martin:2012pe,Chen:2013aj,Huang:2013oya,Chen:2013eea}. Considerable $f_{_\mathrm{NL}}^{(\mathrm{local})}$ can also be generated in multi-field inflation models, such as those implementing the curvaton idea \cite{Mollerach:1989hu,Enqvist:2001zp,Lyth:2001nq,Moroi:2001ct} (see also \cite{Byrnes:2010em} for a review on local non-gaussianity in multi-field inflation). Models with non-BD vacuum \cite{Holman:2007na} and higher-derivative terms \cite{Senatore:2009gt,Bartolo:2010bj} can also induce some potentially observable amount of the general ``folded'' type bispectrum. For inflationary models with non-canonical kinetic terms such as $k$-inflation \cite{ArmendarizPicon:1999rj}, DBI-inflation \cite{Silverstein:2003hf,Alishahiha:2004eh} and their multi-field generalizations such as \cite{Cai:2009hw,Gao:2009at}, or with higher-derivative interactions such as ghost inflation \cite{ArkaniHamed:2003uz} and $G$-inflation \cite{Kobayashi:2010cm,Kobayashi:2011nu}, or in the effective theory of inflation \cite{Cheung:2007st}, the main contribution to the bispectrum comes from the ``equilateral'' and ``orthogonal'' shapes, and can, again, be generically quite large (see \cite{Tsujikawa:2013ila} for a recent summary on non-gaussianities in single-field scalar-tensor models).

Let us now turn to a detailed characterization of the shapes and amplitudes of non-gaussianities across the bouncing phase.  From here on, we set the spatial curvature $\mcl K$ to 1.  Let us first write down the general form of the three distinct matrix elements of $\bm\Pi$ in terms of $P_+$ and $P_-\,$ and in terms of an additional set of ``shape'' functions $K_1$ to $K_6$,
\bea
&& K_1=\sum_{i\ne j}N_iN_j,\qquad\,
K_2=\prod_iN_i,\qquad\qquad\quad K_3=\sum_i\frac{1}{D_i},\nn\\ &&
K_4=\sum_{i,\,j}\frac{1}{D_iD_j},\qquad K_5= \sum_{{\scpt
\begin{array}{c} i,\,j,\,k\\ i\ne j \end{array}}}
\frac{N_iN_j}{D_k},\qquad K_6=\sum_iN_i,
\eea
where we have defined
\bea
N_1&=&\displaystyle \frac{3}{2}k_i^2-4, \quad i=1\\
N_i&=&\displaystyle k_i^2-4, \,\,\,\quad i=2,3\\ D_i&=&\displaystyle
\frac{k_i^2}{2}, \quad\qquad i=1,2,3
\eea
and in terms of the following simplifying combinations
\bea
\frac{F_1}{6}&=&1+D_1\lsb
1-\frac{1}{3}\lb 4P_+-\Xi\rb\rsb+\Xi,\\ F_2&=&27+D_1\lsb
1-\frac{1}{3}\lb 28 P_+-\Xi\rb\rsb+\Xi.
\eea
In the above, $P_{\pm}$ and $\Xi$ are defined in appendix (\ref{Ppm_Pi_def}).
We obtain 
\bea
\Pi_{11}(k_1,k_2,k_3)&=&\frac{1}{3D_1D_2D_3}\Biggl\{ 12
\Upsilon F_2+6N_1\lsb -F_2+\Upsilon\lb -3+\Xi+F_2 \rb\rsb+K_1\lsb
2F_1+12\Upsilon\lb 7-\Xi\rb-72\eta_V\rsb\nn\\ &&+K_2\lb
-\frac{F_1}{\Upsilon}-42+6\Xi+36\frac{\eta_V}{\Upsilon}-12\eta_V\rb-K_3
\lb 12\Upsilon F_2N_1\rb+4\Upsilon K_5 F_1-4\Upsilon K_6 F_1\nn+\\ 
&& K_2K_3\lsb -2F_1+12\Upsilon\lb -7+\Xi\rb\rsb-4\Upsilon K_2K_4 F_1\Biggl\},
\label{eq:Pi11}
\eea
\bea
\Pi_{12}(k_1,k_2,k_3)&=&\frac{1}{D_1D_2D_3}\Biggl(
12\Upsilon\lb-6+F_2\rb+3N_1\lsb 12-2F_2+\Upsilon\lb
-12+F_2\rb\rsb+\nn\\ &&2N_3\lsb F_1+\Upsilon\lb
42-12\Xi-\frac{F_1}{2}\rb-36\eta_V\rsb+N_1N_3\lsb
-\frac{F_1}{\Upsilon}-42+\frac{F_1}{2}+6\Xi+\frac{36\eta_V}{\Upsilon}+
\right.\nn\\ &&\left.\Upsilon\lb
21+\frac{F_1}{8}-3\Xi\rb-30\eta_V\rsb+N_2N_3\lsb F_1+\Upsilon\lb
41-6\Xi+\frac{F_1}{2}\rb-36\eta_V\rsb+\nn\\ &&K_2\lsb
-\frac{F_1}{2\Upsilon}-21-\frac{F_1}{4}+3\Xi+\frac{18\eta_V}{\Upsilon}+
\frac{3}{2}\Upsilon\lb -7-\frac{F_1}{8}+\Xi\rb\rsb+\nn\\
&&K_3\left\{12\Upsilon N_1\lb 6-F_2\rb+4\Upsilon F_1N_3+2\Upsilon
F_1N_2N_3+N_1N_3\left[ -2F_1+\Upsilon\lb
-84+F_1+12\Xi\rb+72\eta_V\right]\right\}-\nn\\
&&4\Upsilon K_4 N_1N_3F_1+K_2K_3\lsb -F_1+6\Upsilon\lb
-8+\Xi\rb+36\eta_V\rsb-2\Upsilon F_1K_2K_4\Biggl),
\label{eq:Pi12}
\eea
and
\bea
\Pi_{22}(k_1,k_2,k_3)&=&\frac{1}{6D_1D_2D_3}\Biggl\{ 24\Upsilon\lb 6
+F_2\rb+12N_1\lsb 6+\Upsilon\lb -3+\Xi\rb\rsb +N_2N_3\lsb
F_1+\Upsilon\lb F_1+42-6\Xi\rb-36\eta_V\rsb\nn\\ &&+K_2\lsb
-\frac{F_1}{2\Upsilon}-21+\frac{F_1}{2}-12N_1+3\Xi+18\frac{\eta_V}{
\Upsilon}+\Upsilon\lb -21-\frac{F_1}{2}+3\Xi\rb+12\eta_V\rsb\nn\\ &&
+2\Upsilon\,K_3\lsb F_1N_2N_3+12\lb 6-F_2\rb\rsb +K_2K_3\lsb
-F_1+\Upsilon\lb -42-F_1+6\Xi\rb+36\eta_V\rsb\nn\\ &&-2\Upsilon K_2K_4 F_1 \Biggl\}.
\label{eq:Pi22}
\eea
To understand which basic shapes produce the largest amount of non-gaussianities, it is then sufficient to understand the behaviour of the functions $P_+$, $P_-$ and $K_1$ to $K_6$.  The functions $P_+$ and $P_-$ peak in the squeezed and folded configurations respectively. The shape functions $K_1$, $K_2$ and $K_6$ peak in the equilateral configuration and at large $k$'s.  We can rewrite $K_3$ and $K_4$ as
\bea
K_3&=&\frac{1}{D_1}\lb
1+\frac{D_1}{D_2}+\frac{D_1}{D_3}\rb,\\ K_4&=&\frac{1}{D_1^2}\lsb 1+\lb
\frac{D_1}{D_2}\rb^2+\lb
\frac{D_1}{D_3}\rb^2+\frac{D_1}{D_2}+\frac{D_1}{D_3}+\frac{D_1}{D_2}
\frac{D_1}{D_3}\rsb\,,
\eea
and, keeping in mind the constraints imposed by
the triangle inequalities of the $n_i$'s, we find that both
$K_3$ and $K_4$ peak in the equilateral shape and on large scales.
Finally, the shape function $K_5$ can be written as
\bea
K_5=
\sum_{{\scpt \begin{array}{c} i,\,j,\,k\\ i\ne j\ne k \end{array}}}
\frac{N_iN_j}{D_k} +\sum_{{\scpt \begin{array}{c} i,\,j\\ i\ne j
\end{array}}} \frac{N_iN_j}{D_j} 
\eea
In this expression, only the first term is of any real
interest and is found to peak in the squeezed configuration.\\

It is worth noting that $P_-$ only appears through the presence of $\Xi$ in the expressions (\ref{eq:Pi11}) to (\ref{eq:Pi22}).  The folded shape is therefore suppressed by an additional factor $1/(k^2-3\mcl K)$ relative to the squeezed and equilateral shapes.  This analysis in terms of shape functions thus suggests that on scales of cosmological interest today the non-gaussianities produced in a symmetric bouncing phase are largest for the set of (linear or bilinear) combinations of two basic shapes, namely the equilateral and squeezed shapes.  Furthermore, it can be checked that the inclusion of some level of asymmetry in the cosmological evolution (\ie $\varepsilon_V$ non-zero) contributes additional non-gaussianities whose leading terms are proportional to the ratio $\varepsilon_V/\Upsilon^{3/2}$.  The asymmetry thus contributes only subleading terms (in powers of the set of parameters $\Upsilon$ to $\eta_V$) relative to the leading symmetric contribution proportional to $1/\Upsilon$.  Therefore, any asymmetry can {\it a priori} only further enhance the level of non-gaussianity at next-to-leading order.  In fact, unless the ratio $\varepsilon_V/\Upsilon^{3/2}$ is fine-tuned, and unless fortuitous cancellations occur, and especially because the $k$-dependences of symmetric and asymmetric non-gaussian terms differ, the amount of non-gaussianity produced by the bouncing phase cannot possibly be significantly reduced by including some level of asymmetry in the bounce.
In the next three sections, we evaluate the matrix elements of $\bm \Pi$ in the equilateral, squeezed, and local shapes.

\subsection{Equilateral configuration}
\label{sub:equi}

In the equilateral case,
$k_1=k_2=k_3=k$ and all six contributions in the numerator of the
bispectrum are equal, while all three terms in the denominator are also
equal.  We thus have
\begin{equation}
f_{_\mathrm{NL}}^{\mrm{(equil)}}=\frac{5}{3}\frac{v^{T}\left(k,\eta\right)
\bm{P}\left(k\right)\bm{\Pi}\left(k,k,k;\eta\right)\bm{P}\left(k\right)v
\left(k,\eta\right)}{v^{T}\left(k,\eta\right)\bm{P}\left(k\right)v\left(k
,\eta\right)v^{T}\left(k,\eta\right)\bm{P}\left(k\right)v\left(k,\eta
\right)}.
\end{equation}
As mentioned previously, there are only four matrix entries
that describe the amplification of non gaussiannities across the bounce,
but the off-diagonal terms in $\bm\Pi$ are equal. This leaves only three
distinct contributions to the matrix elements of $\bm \Pi$.  Here, we
present the resulting expressions as series expansions in powers of $k$
and up to first order in $\{\Upsilon,\eta_V\}$:
\bea
\Pi_{11}(k,k,k) & \simeq &
\frac{216}{\Upsilon}k^2-\frac{1704}{\Upsilon}+1008+\frac{864\eta_V}{
\Upsilon}-288\eta_V+\nn\\ 
&&\lb
\frac{3560}{\Upsilon}-6872-\frac{5760\eta_V}{\Upsilon}+4680\Upsilon+7680
\eta_V\rb k^{-2}+\mcl O(k^{-4}),\\ \Pi_{12}(k,k,k) & \simeq & \left(
54+\frac{108}{\Upsilon}+\frac{81\Upsilon}{2}\right) k^2
-\frac{1068}{\Upsilon}+186+\frac{432\eta_V}{\Upsilon}-\frac{81\Upsilon}{
2}+72\eta_V+\nn\\
&&\lb
\frac{3052}{\Upsilon}-3938-\frac{3744\eta_V}{\Upsilon}+\frac{3017
\Upsilon}{2}+3120\eta_V\rb k^{-2}+\mcl O(k^{-4}),\\ \Pi_{22}(k,k,k) &
\simeq & \left( \frac{54}{\Upsilon}+54+54\Upsilon\right)
k^2-\frac{642}{\Upsilon}-174+\frac{216\eta_V}{\Upsilon}-174\Upsilon+144
\eta_V+\nn\\ 
&&\lb
\frac{2378}{\Upsilon}-1458-\frac{2304\eta_V}{\Upsilon}+342\Upsilon+768
\eta_V\rb k^{-2}+\mcl O(k^{-4}).
\eea
The parameter $\Upsilon$ appearing
in the denominator of some of the terms in these matrix element
corresponds to the square of the scalar field $\phi$'s first time
derivative at the bounce point and is related to the characteristic
timescale of the bounce, $\eta_c$.  As $\Upsilon\rightarrow 0$, $\eta_c$
approaches unity, and the bounce approaches an exact de Sitter bounce.  The leading terms in the matrix elements $\Pi_{ab}$, $(a,b=1,2)$, scale as $k^2/\Upsilon$, thus peaking on small scales and for $\Upsilon\ll 1$.
\subsection{Squeezed limit}
\label{sub:local}
The local shape is obtained by demanding $k_2=k_3=k$ and $k_1=p$,
with $\sqrt{8}\leq p\ll k$ (the lower bound being due to the fact that the
values $n=0,1$ correspond to gauge modes). The corresponding nonlinear
parameter $f_{_\mathrm{NL}}$ now reads 
{\small
\begin{equation}
f_{_\mathrm{NL}}^{\mathrm{(sq)}}=\frac{5}{3}\frac{v^{T}(p)\bm P\lb
p\rb\bm{\Pi}\lb k,p,k\rb\bm{P}(k)v(k)+v^{T}(k)\bm P\lb k\rb\bm{\Pi}\lb
k,k,p\rb\bm{P}(p)v(p)+v^{T}(k)\bm P\lb k\rb\bm{\Pi}\lb
p,k,k\rb\bm{P}(k)v(k)}{2v^{T}(p)\bm{P}(p)v(p)v^{T}(k)\bm{P}(k)v(k)+v^{T}
(k)\bm{P}(k)v(k)v^{T}(k)P(k)v(k)}.
\end{equation}
}
As explained in Section \ref{sec:bispectrum}, there are a total of seven distinct contributions to the nonlinear parameter $f_{_\mathrm{NL}}$, given by the following matrix elements of $\bm \Pi$,
\bea
\Pi_{11}(p,k,k) & \simeq & \frac{k^2}{p^2-3}\lsb
-\frac{576}{\Upsilon}+448+\frac{96p^2}{\Upsilon}+\lb
\frac{1408}{3\Upsilon}-\frac{6016}{3}+1536\Upsilon\rb p^{-2}+\right.\\
&&\left.\lb
\frac{1024}{\Upsilon}+\frac{4096}{3}-\frac{25600\Upsilon}{3}\rb
p^{-4}+\lb 4096+\frac{16384\Upsilon}{3}\rb p^{-6}+16384\Upsilon
p^{-8}+\mcl O(k^{-2})\rsb,\nn\\ \Pi_{11}(k,p,k) & \simeq &
-\frac{192}{\Upsilon}-192+\frac{288\eta_V}{\Upsilon}-96\eta_V+\frac{96p^
2}{\Upsilon}+\nn\\ &&\lb
192-\frac{576\eta_V}{\Upsilon}-192\Upsilon+768\eta_V\rb p^{-2}+\lb
768\Upsilon-2304\eta_V\rb p^{-4}+\mcl O(k^{-2}),
\eea
and
\bea
\Pi_{22}(p,k,k) &
\simeq & \frac{k^2}{p^2-3}\lsb -\frac{144}{\Upsilon}-32-32\Upsilon +\lb
\frac{24}{\Upsilon}+24+24\Upsilon\rb p^2+\lb
\frac{352}{3\Upsilon}-384\rb p^{-2}+\nn\right.\\ &&\left.\lb
\frac{256}{\Upsilon}+\frac{1792}{3}-1536\Upsilon\rb p^{-4}+\lb
1024+\frac{7168\Upsilon}{3}\rb p^{-6}+4096\Upsilon p^{-8}+\mcl
(k^{-2})\rsb,\\
\Pi_{22}(k,p,k) & \simeq &
-\frac{96}{\Upsilon}-96+\frac{72\eta_V}{\Upsilon}-96\Upsilon+48\eta_V+
\lb \frac{24}{\Upsilon}+24+24\Upsilon\rb p^2+\nn\\ &&\lb
192-\frac{288\eta_V}{\Upsilon}+96\eta_V\rb p^{-2}+\lb
768\Upsilon-1152\eta_V\rb p^{-4}+\mcl O(k^{-2}),
\eea
for the diagonal terms, the non-diagonal ones reading
\bea
\Pi_{12}(p,k,k) &
\simeq & \frac{k^2}{p^2-3}\lsb -\frac{288}{\Upsilon}+80+4\Upsilon+\lb
\frac{48}{\Upsilon}+24+18\Upsilon\rb p^2+\lb
\frac{704}{2\Upsilon}-\frac{2656}{3}+\frac{1064\Upsilon}{3}\rb
p^{-2}+\right.\nn\\ &&\left. \lb
\frac{512}{\Upsilon}+\frac{2816}{3}-\frac{11200\Upsilon}{3}\rb
p^{-4}+\lb 2048+\frac{11264\Upsilon}{3}\rb p^{-6}+8192\Upsilon
p^{-8}+\mcl O(k^{-2})\rsb,\\ \Pi_{12}(k,p,k) & \simeq &
-\frac{96}{\Upsilon}-144+\frac{144\eta_V}{\Upsilon}-84\Upsilon+24\eta_V+
\lb \frac{48}{\Upsilon}+24+18\Upsilon\rb p^2+\nn\\ &&\lb
192-\frac{288\eta_V}{\Upsilon}-96\Upsilon+240\eta_V\rb p^{-2}+\lb
768\Upsilon-1152\eta_V\rb p^{-4}+\mcl O(k^{-2}),\\
\Pi_{12}(k,k,p) & \simeq &
-\frac{192}{\Upsilon}-96+\frac{144\eta_V}{\Upsilon}-72\Upsilon+24\eta_V+
\lb \frac{48}{\Upsilon}+24+18\Upsilon\rb p^2+\nn\\ &&\lb
192-\frac{576\eta_V}{\Upsilon}-96\Upsilon+480\eta_V\rb p^{-2}+\lb
768\Upsilon-2304\eta_V\rb p^{-4}+\mcl O(k^{-2}).
\eea
The leading contribution in the squeezed configuration scales as $k^2/(\Upsilon p^2)$.  As in the case of the equilateral configuration, local-type non-gaussianities will be enhanced on small scales and for a bounce approaching a pure de Sitter bounce.

\subsection{Folded configuration}
\label{sub:folded}
The folded case is defined by the requirement $k_1=2k$ and $k_2=k_3=k$
on the Fourier modes. In this case, we have

{\small
\begin{equation}
f_{_\mathrm{NL}}^{\mrm{(fold)}}=\frac{5}{3}\frac{v^{T}(2k)\bm P\lb
2k\rb\bm{\Pi}\lb k,2k,k\rb\bm{P}(k)v(k)+v^{T}(k)\bm P\lb k\rb\bm{\Pi}\lb
k,k,2k\rb\bm{P}(2k)v(2k)+v^{T}(k)\bm P\lb k\rb\bm{\Pi}\lb
2k,k,k\rb\bm{P}(k)v(k)}{2v^{T}(2k)\bm{P}(2k)v(2k)v^{T}(k)\bm{P}(k)v(k)+v
^{T}(k)\bm{P}(k)v(k)v^{T}(k)P(k)v(k)}.
\end{equation}
}

In the observable range of lengthscales, the relevant matrix entries are
\bea
\Pi_{11}(2k,k,k) & \simeq &
-\frac{64}{\Upsilon}-288+\frac{288\eta_V}{\Upsilon}-96\eta_V+\nn\\ &&\lb
\frac{824}{3\Upsilon}+\frac{4432}{3}-\frac{1344\eta_V}{\Upsilon}-1728
\Upsilon+1792\eta_V\rb k^{-2}+\mcl O(k^{-4}),
\\
\Pi_{11}(k,2k,k) & \simeq
& \frac{512}{\Upsilon}-1152+\frac{288\eta_V}{\Upsilon}-96\eta_V+\nn\\
&&\lb
-\frac{5632}{3\Upsilon}+\frac{21904}{3}-\frac{1488\eta_V}{\Upsilon}-6192
\Upsilon+1984\eta_V\rb k^{-2}+\mcl O(k^{-4}),
\eea
and
\bea
\Pi_{22}(2k,k,k)
& \simeq &
-\frac{16}{\Upsilon}-88+\frac{72\eta_V}{\Upsilon}-88\Upsilon+48\eta_V+
\nn\\ && \lb
\frac{398}{3\Upsilon}+1014-\frac{624\eta_V}{\Upsilon}+6\Upsilon+208
\eta_V\rb k^{-2}+\mcl O(k^{-4}),\\ \Pi_{22}(k,2k,k) & \simeq &
\frac{128}{\Upsilon}-160+\frac{72\eta_V}{\Upsilon}-160\Upsilon+48\eta_V+
\nn\\ &&\lb
-\frac{2368}{3\Upsilon}+2256-\frac{552\eta_V}{\Upsilon}-192\Upsilon+184
\eta_V\rb k^{-2}+\mcl O(k^{-4}),
\eea
for the independent diagonal terms, and
\bea
\Pi_{12}(2k,k,k) & \simeq
&
-\frac{32}{\Upsilon}-160+\frac{140\eta_V}{\Upsilon}-84\Upsilon+24\eta_V+
\nn\\ && \lb
\frac{604}{3\Upsilon}+\frac{3766}{3}-\frac{960\eta_V}{\Upsilon}-\frac{
4147\Upsilon}{6}+800\eta_V\rb k^{-2}+\mcl O(k^{-4}),
\\
\Pi_{12}(k,2k,k) &
\simeq &
\frac{256}{\Upsilon}-448+\frac{144\eta_V}{\Upsilon}-192\Upsilon+24\eta_V
+\nn\\ &&\lb
-\frac{4352}{3\Upsilon}+\frac{14128}{3}-\frac{1032\eta_V}{\Upsilon}-
\frac{6584\Upsilon}{3}+860\eta_V\rb k^{-2}+\mcl O(k^{-4}),
\\
\Pi_{12}(k,k,2k) & \simeq &
\frac{256}{\Upsilon}-448+\frac{144\eta_V}{\Upsilon}-192\Upsilon+24\eta_V
+\nn\\ &&\lb
-\frac{3200}{3\Upsilon}+\frac{10960}{3}-\frac{816\eta_V}{\Upsilon}-\frac
{5432\Upsilon}{3}+680\eta_V\rb k^{-2}+\mcl O(k^{-4}),
\eea
for the non-diagonal terms.

In the folded configuration, the matrix elements of $\bm \Pi$ are largely $k$-independent for wavenumbers in the observable range, and proportional to $1/\Upsilon$. As anticipated in Section \ref{sec:shapefunctions}, folded-type non-gaussianities are thus subdominant relative to equilateral and local type non-gaussianities.  This is particularly true on scales of observable interest today, for which the wavenumber $k\gg 1$.

\section{Conclusion}

Bouncing cosmologies have been presented as plausible alternatives to
the inflationary paradigm \cite{Brandenberger:2012zb} and should thus be
confronted to all existing data. To date, the power spectrum has been
calculated in many models, and, in some cases, has been found to be,
{\it potentially}, in agreement with observations. The computation of the three-point function provides a measure of the amplitude and shape of the non-gaussian signal that might be present in the statistics of cosmological perturbations and provides another powerful tool to discriminate between viable and non-viable cosmological models given that the level of non-gaussianity in the CMB has been severely constrained by the recently released Planck data.  The calculation of the three-point function has only received limited attention in the context of bouncing cosmologies, and in particular, the explicit calculation of the production of non-gaussianities by the bouncing phase itself has been lacking.  In the present work, we give a detailed computation of the amplitude and shape of non-gaussianities produced in a bouncing scenario obtained within the context of GR and with a single scalar field having a standard kinetic term.  The bounce is obtained by assuming the positivity of the spatial sections of the background FLRW metric, an assumption which is by no means incompatible with the currently observed smallness of $\Omega_{\mcl K}$.

By expanding the scale factor, the scalar field $\phi$ and the potential $V(\phi)$ in powers of the conformal time around the bounce point, and by considering a set of generic gaussian initial conditions for the first order cosmological perturbations on a spatial hypersurface at some infinitesimal time $\eta_-$ prior to the bounce, we have computed the exact evolution of the second order perturbation, which is sourced by the first order perturbation and amplified by the non-trivial background evolution at the bounce.  Through the analysis of the matrix $\bm\Pi$, we have also provided estimates of the shapes and amount of non-gaussianity produced by the bouncing phase.  On the one hand, the shapes of the non-gaussianity produced depend on the existence of two mode functions (as opposed to the non-gaussianity produced in inflation, which depends on the growing mode only, the decaying mode being discarded) and on the structure of the background spacetime, in this instance an FLRW spacetime with closed spatial sections.  On the other hand, the amplitude of the non-gaussian signal is parameterized in terms of a set of small parameters $\Upsilon$ to $\eta_V$ which characterize the deviation of the bouncing cosmology from a pure de Sitter bounce obtainable with positive spatial curvature and a positive cosmological constant.

We concluded this work by computing analytic formulae for three of the most widely estimated shapes, namely the equilateral, squeezed and folded configurations. In an expansion of the three-point function in powers of the parameters $\Upsilon$ to $\eta_V$, the largest non-gaussian terms were found to be inversely proportional to the small parameter $\Upsilon$ which encodes the deviation of the characteristic bouncing timescale from the one of a purely de Sitter bounce.  At subleading orders, there also exists some level of non-gaussianities proportional to the parameters that encode symmetric and asymmetric deviations away from the shape of the purely de Sitter bounce.  Interestingly, both the squeezed and equilateral shapes contribute terms to the matrix elements of $\Pi$ proportional to $(k/p)^2$ and $k^2$ respectively, while the folded configuration contributes terms proportional to $1/k^2$. 

Although the determination of the dominant shapes of non-gaussianities produced and of the corresponding amplitudes was performed in the simpler case of a symmetric bouncing phase, it was checked that any asymmetry in the shape of the bounce produces subleading contributions to the level of non-gaussianity.  Any asymmetry can thus {\it a priori} only further enhance the amount of non-gaussianity generated by a curvature-dominated bounce.  

Having established the general formalism to calculate the non-gaussianities induced in the gravitational potential by a bouncing phase, there remains to evaluate the non-linearity parameter $f_{\mrm{NL}}$ itself. This is the subject of a forthcoming paper~\cite{GaoLilleyPeter14}.  This task introduces some level of model-dependence, since in order to provide precise numerical estimates of the parameter $f_{_\mrm{NL}}$, one has to make assumptions on the values of the parameters 
$\Upsilon$, $\varepsilon_V$ and $\eta_V$ and on the precise form of
the primordial power spectral matrix $\bm P$ introduced in Section \ref{sub:initcond}.

\acknowledgments X.G. was supported by ANR (Agence Nationale de la Recherche) grant STR-COSMO ANR-09-BLAN-0157-01 and JSPS Grant-in-Aid for Scientific Research No. 25287054. M.L. acknowledges the Laboratoire AstroParticule \& Cosmologie and Universit\'e Denis Diderot-Paris 7 where this work was initiated.  M.L. also acknowledges the University of Cape Town and is currently supported by a South African NRF research grant.

\bibliography{references}

\appendix

\section{Second order scalar perturbation equation}
\label{EinsteinEqs}

We use the FLRW metric in Poisson gauge and consider
only scalar perturbations, \ie we consider the perturbed metric
(\ref{metric_pert_poisson})
\bea \dd s^{2}=g_{\mu\nu}\dd x^{\mu}\dd
x^{\nu}=a^{2}\left(-\ex^{2\Phi}\dd\eta^{2} +\ex^{-2\Psi}\gamma_{ij}\dd
x^{i}\dd x^{j}\right),
\label{metric_phy_newton_gen}
\ea
from which we
derive the various relevant tensor below.

\subsection{Einstein tensor}
The components of the Einstein tensor $G_{\mu\nu}\equiv R_{\mu\nu} -
\frac12 g_{\mu\nu} R$ read
\begin{eqnarray}
G_{00} & = &
3\left(\mathcal{H}-\Psi'\right)^{2}+\ex^{2\left(\Phi+\Psi\right)}\left[2
\bar{\nabla}^{2}\Psi-\left(\bar{\nabla}_{i}\Psi\right)^{2}+3K\right],
\label{G_00}\\
G_{0i} & = & 2\left(\mathcal{H}-\Psi'\right)\bar{\nabla}_{i}\Phi+2\bar{\nabla}_{i}
\Psi',\label{G_0i}\\
G_{ij} & = & \gamma_{ij}\Big\{
\ex^{-2\left(\Phi+\Psi\right)}\left[-2\mathcal{H}'-\left(\mathcal{H}-2
\Phi'-3\Psi'\right)\left(\mathcal{H}-\Psi'\right)+2\Psi''\right]+
\bar{\nabla}^{2}\left(\Phi-\Psi\right)+\left(\bar{\nabla}_{k}\Phi\right)^{2}-K
\Big\}\nonumber \\
&  & -\bar{\nabla}_{i}\bar{\nabla}_{j}\left(\Phi-\Psi\right)-\bar{\nabla}_{i}
\Phi\bar{\nabla}_{j}\Phi+\bar{\nabla}_{i}\Psi\bar{\nabla}_{j}\Psi-
\bar{\nabla}_{i}\Phi\bar{\nabla}_{j}\Psi-\bar{\nabla}_{j}\Phi\bar{\nabla}_{i}
\Psi\,,\label{G_ij}
\end{eqnarray}
while the trace and trace-free parts of $G_{ij}$ are 
\bea
\mathrm{tr}\,\lb G_{ij}\rb=\gamma^{ij}G_{ij}&=&3\ex^{-2\left(\Phi+\Psi\right)}\left[-2\mathcal{H}'-\left(\mathcal{H}-2\Phi'-3\Psi'\right)\left(\mathcal{H}-\Psi'\right)+2\Psi''\right]\nonumber\\
&& +2\bar{\nabla}^{2}\left(\Phi-\Psi\right)+2\left(\bar{\nabla}_{i}\Phi
\right)^{2}+\left(\bar{\nabla}_{i}\Psi\right)^{2}-2\bar{\nabla}_{i}\Phi
\bar{\nabla}^{i}\Psi-3K\,, \label{G_ij_tr}\\
G_{ij}^{(\mathrm{tf})}=G_{ij}-\frac{1}{3}\gamma_{ij}\mathrm{tr}\,\lb G_{ij}\rb &=&
\frac{1}{3}\gamma_{ij}\left[\bar{\nabla}^{2}\left(\Phi-\Psi\right)+\left
(\bar{\nabla}_{k}\Phi\right)^{2}-\left(\bar{\nabla}_{k}\Psi\right)^{2}+2
\bar{\nabla}_{k}\Phi\bar{\nabla}^{k}\Psi\right]\nn\\ 
&&-\bar{\nabla}_{i}\bar{\nabla}_{j}\left(\Phi-\Psi\right)-\bar{\nabla}_{i}
\Phi\bar{\nabla}_{j}\Phi+\bar{\nabla}_{i}\Psi\bar{\nabla}_{j}\Psi-
\bar{\nabla}_{i}\Phi\bar{\nabla}_{j}\Psi-\bar{\nabla}_{j}\Phi\bar{\nabla}_{i}
\Psi. \label{G_ij_tf}
\end{eqnarray} 
The components of $G_{\mu}^{\nu}$ are given by 
\bea G_{0}^{0} & = &
-3\frac{\ex^{-2\Phi}}{a^{2}}\left(\mathcal{H}-\Psi'\right)^{2}-\frac{\ex
^{2\Psi}}{a^{2}}\left[2\bar{\nabla}^{2}\Psi-\left(\bar{\nabla}_{i}\Psi
\right)^{2}+3K\right],\label{G^0_0}\\ 
G_{i}^{0} & = &
-2\frac{\ex^{-2\Phi}}{a^{2}}\left[\left(\mathcal{H}-\Psi'\right)
\bar{\nabla}_{i}\Phi+\bar{\nabla}_{i}\Psi'\right],\label{G^0_i}\\ 
G_{0}^{i} &= & 2\frac{\ex^{2\Psi}}{a^{2}}\left[\left(\mathcal{H}-\Psi'\right)
\bar{\nabla}^{i}\Phi+\bar{\nabla}^{i}\Psi'\right],\label{G^i_0}\\ 
G_{j}^{i} &= & \delta_{j}^{i}\frac{1}{a^{2}}\left\{
\ex^{-2\Phi}\left[-2\mathcal{H}'-\left(\mathcal{H}-2\Phi'-3\Psi'\right)
\left(\mathcal{H}-\Psi'\right)+2\Psi''\right]+\ex^{2\Psi}\left[
\bar{\nabla}^{2}\left(\Phi-\Psi\right)+\left(\bar{\nabla}_{i}\Phi\right)^{2}-K
\right]\right\} \nonumber\\ 
&  &
-\frac{\ex^{2\Psi}}{a^{2}}\left[\bar{\nabla}^{i}\bar{\nabla}_{j}
\left(\Phi-\Psi\right)+\bar{\nabla}^{i}\Phi\bar{\nabla}_{j}\Phi-\bar{\nabla}^{i
}\Psi\bar{\nabla}_{j}\Psi+\bar{\nabla}^{i}\Phi\bar{\nabla}_{j}\Psi+\bar{
\nabla}_{j}\Phi\bar{\nabla}^{i}\Psi\right]\,,
\label{G^i_j} 
\eea while the trace and trace-free parts of $G_{j}^{i}$ are
\bea
\mathrm{tr}\,\lb
G_{j}^{i}\rb=\delta_{i}^{j}G_{j}^{i} &=&
3\frac{\ex^{-2\Phi}}{a^{2}}\left[-2\mathcal{H}'-\left(\mathcal{H}-2\Phi'
-3\Psi'\right)\left(\mathcal{H}-\Psi'\right)+2\Psi''\right]\nonumber \\
&  &
+\frac{\ex^{2\Psi}}{a^{2}}\left[2\bar{\nabla}^{2}\left(\Phi-\Psi\right)+
2\left(\bar{\nabla}_{i}\Phi\right)^{2}+\left(\bar{\nabla}_{i}\Psi\right)
^{2}-2\bar{\nabla}_{i}\Phi\bar{\nabla}^{i}\Psi-3K\right],
\label{G^i_j_tr}\\
G_{j}^{\mathrm{(tf)}\,i}=G_{j}^{i}-\frac{1}{3}\delta_{j}^{i}\mathrm{tr}\,\lb G_{j}^{i}\rb
&=&\frac{\ex^{2\Psi}}{a^{2}}\Big\{\frac{1}{3}\delta_{j}^{i}\left[
\bar{\nabla}^{2}\left(\Phi-\Psi\right)+\left(\bar{\nabla}_{k}\Phi\right)^{2}-
\left(\bar{\nabla}_{k}\Psi\right)^{2}+2\bar{\nabla}_{k}\Phi\bar{\nabla}^{
k}\Psi\right]\nonumber \\ &  &
-\bar{\nabla}^{i}\bar{\nabla}_{j}\left(\Phi-\Psi\right)-\bar{\nabla}^{i}
\Phi\bar{\nabla}_{j}\Phi+\bar{\nabla}^{i}\Psi\bar{\nabla}_{j}\Psi-
\bar{\nabla}^{i}\Phi\bar{\nabla}_{j}\Psi-\bar{\nabla}_{j}\Phi\bar{\nabla}^{i}
\Psi\Big\}.\label{G^i_j_tf} 
\end{eqnarray}

\subsection{Energy-momentum tensor}

Let us consider the simple single scalar field model,
\begin{equation}
\mathcal{L}=X-V\left(\phi\right),
\end{equation}
with $X\equiv-\frac{1}{2}\partial_{\mu}\phi\partial^{\mu}\phi$. It is easy to
evaluate the energy-momentum tensor
$T_{\mu\nu}=\left(X-V\right)g_{\mu\nu}+\partial_{\mu}\phi\partial_{\nu}
\phi$, where \begin{equation}X\equiv
-\frac{1}{2}g^{\mu\nu}\partial_{\mu}\phi\partial_{\nu}\phi=\frac{\ex^{-2
\Phi}}{2a^{2}}\left(\phi'\right)^{2}-\frac{\ex^{2\Psi}}{2a^{2}}\left(
\bar{\nabla}_{i}\phi\right)^{2}, \label{X_expl}
\end{equation}
with
$\bar{\nabla}^{i}\phi\equiv\gamma^{ij}\bar{\nabla}_{j}\phi$ and
$\left(\bar{\nabla}_{i}\phi\right)^{2}=\gamma^{ij}\bar{\nabla}_{i}\phi
\bar{\nabla}_{j}\phi$. 
The components of $T_{\mu\nu}$ thus read \bea
T_{00} & = &
\frac{1}{2}\left(\phi'\right)^{2}+\frac{\ex^{2\left(\Phi+\Psi\right)}}{2
}\left(\bar{\nabla}_{i}\phi\right)^{2}+a^{2}\ex^{2\Phi}V\left(\phi\right
),\label{T_00}\\ T_{0i} & = & \phi'\bar{\nabla}_{i}\phi,\label{T_0i}\\
T_{ij} & = &
\lsb\frac{\ex^{-2\Phi}}{2a^{2}}\left(\phi'\right)^{2}-\frac{\ex^{2\Psi}}
{2a^{2}}\left(\bar{\nabla}_{k}\phi\right)^{2}-V\left(\phi\right)\rsb
a^{2}\ex^{-2\Psi}\gamma_{ij}+\bar{\nabla}_{i}\phi\bar{\nabla}_{j}\phi,
\label{T_ij} \eea
while the trace and trace-free parts of $T_{ij}$ are
\bea \mathrm{tr}\,\lb T_{ij}\rb = \gamma^{ij}T_{ij}
&=&\frac{3}{2}\ex^{-2\left(\Phi+\Psi\right)}\left(\phi'\right)^{2}-\frac
{1}{2}\left(\bar{\nabla}_{i}\phi\right)^{2}-3a^{2}\ex^{-2\Psi}V\left(
\phi\right), \label{T_ij_tr}\\ T_{ij}^{\mathrm{(tf)}} = 
T_{ij}-\frac{1}{3}\gamma_{ij}\lsb\mathrm{tr}\,\lb T_{ij}\rb\rsb & = &
\bar{\nabla}_{i}\phi\bar{\nabla}_{j}\phi-\frac{1}{3}\gamma_{ij}\lb\bar{
\nabla}_{k}\phi\rb^{2}. \label{T_ij_tf} \eea
The components of
$T_{\mu}^{\nu}$ read \bea T_{0}^{0} & = &
-\frac{\ex^{-2\Phi}}{2a^{2}}\left(\phi'\right)^{2}-\frac{\ex^{2\Psi}}{2a
^{2}}\left(\bar{\nabla}_{i}\phi\right)-V\left(\phi\right),
\label{T^0_0}\\ T_{i}^{0} & = &
-\frac{\ex^{-2\Phi}}{a^{2}}\phi'\bar{\nabla}_{i}\phi, \label{T^0_i}\\
T_{0}^{i} & = & \frac{\ex^{2\Psi}}{a^{2}}\phi'\bar{\nabla}^{i}\phi,
\label{T^i_0}\\ T_{j}^{i} & = &
\lsb\frac{\ex^{-2\Phi}}{2a^{2}}\left(\phi'\right)^{2}-\frac{\ex^{2\Psi}}
{2a^{2}}\left(\bar{\nabla}_{k}\phi\right)^{2}-V\left(\phi\right)\rsb
\delta_{j}^{i}+\frac{\ex^{2\Psi}}{a^{2}}\bar{\nabla}^{i}\phi\bar{\nabla}_
{j}\phi, \label{T^i_j} \eea
while the trace and trace-free parts of $T_{j}^{i}$ are
\bea \mathrm{tr}\,\lb T_{j}^{i}\rb =
\delta_{i}^{j}T_{j}^{i} &=&
\frac{3}{2}\frac{\ex^{-2\Phi}}{a^{2}}\left(\phi'\right)^{2}-\frac{\ex^{2
\Psi}}{2a^{2}}\left(\bar{\nabla}_{i}\phi\right)^{2}-3V\left(\phi\right),
\label{T^i_j_tr}\\ T_{j}^{\mathrm{(tf)}\,i} =
T_{j}^{i}-\frac{1}{3}\delta_{j}^{i}\mathrm{tr}\,\lb T_{j}^{i}\rb &=&
\frac{\ex^{2\Psi}}{a^{2}}\left[\bar{\nabla}^{i}\phi\bar{\nabla}_{j}\phi-
\frac{1}{3}\delta_{j}^{i}\left(\bar{\nabla}_{k}\phi\right)^{2}\right].
\label{T^i_j_tf} \eea \subsection{Einstein equations} In this section,
we expand the Einstein equations up to second order using \begin{equation}X(\eta,\bm
x)= X_{(1)}(\eta,\bm x)+\frac{1}{2}X_{(2)}(\eta,\bm x)+\dots \end{equation}where
$X$ stands for either $\Phi$, $\Psi$ or $\phi$.
\noindent The $G_i^0=T_i^0$ equation reads \begin{equation}2\left(\mathcal{H}-\Psi'\right)\bar{\nabla}_{i}\Phi+2\bar{\nabla}_{i}
\Psi'=\phi'\bar{\nabla}_{i}\phi, \label{EE_0i} \end{equation}and can be used to
express the perturbations $\phi_{(i)}$ in terms of the $\Phi_{(i)}$ and
$\Psi_{(i)}$ ($i=1,2$).  It yields, at first and second order
respectively, \bea
&&2\bar{\nabla}_{i}\lb\Psi_{(1)}'+\mathcal{H}\Phi_{(1)}\rb=\bar{\phi}'
\bar{\nabla}_{i}\phi_{(1)},\label{EE_0i_1st}\\
&&\bar{\nabla}_{i}\lb\Psi_{(2)}'+\mathcal{H}\Phi_{(2)}\rb-2\Psi_{(1)}'
\bar{\nabla}_{i}\Phi_{(1)}=\frac{1}{2}\bar{\phi}'\bar{\nabla}_{i}\phi_{(
2)}+\phi_{(1)}'\bar{\nabla}_{i}\phi_{(1)}.\label{EE_0i_2nd} \eea The
first order equation can easily be solved using the gradient theorem and
assuming the fields vanish at infinity to yield \begin{equation}\phi_{(1)}=\frac{2}{\bar{\phi}'}\lb\Psi_{(1)}'+\mathcal{H}\Phi_{(1)}\rb,
\label{phi_1st_sol} \end{equation}while acting on both sides of (\ref{EE_0i_2nd})
with $\bar{\nabla}^{i}$ yields \begin{equation}\bar{\nabla}^{2}\left(\Psi_{(2)}'+\mathcal{H}\Phi_{(2)}\right)-2\bar{
\nabla}^{i}\left(\Psi_{(1)}'\bar{\nabla}_{i}\Phi_{(1)}\right)=\frac{1}{2}
\bar{\phi}'\bar{\nabla}^{2}\phi_{(2)}+\bar{\nabla}^{i}\left(\phi_{(1)}'
\bar{\nabla}_{i}\phi_{(1)}\right), \end{equation}from which we obtain \begin{equation}\phi_{(2)}=\frac{2}{\bar{\phi}'}\left[\Psi_{(2)}'+\mathcal{H}\Phi_{(2)}-
\bar{\nabla}^{-2}\bar{\nabla}^{i}\left(2\Psi_{(1)}'\bar{\nabla}_{i}\Phi_
{(1)}+\phi_{(1)}'\bar{\nabla}_{i}\phi_{(1)}\right)\right].
\label{phi_2nd_sol_ori} \ee
The $G_j^{(\mrm{tf})\,i}=T_j^{(\mrm{tf})\,i}$ equation reads \bea
&&\frac{1}{3}\delta_{j}^{i}\left[\bar{\nabla}^{2}\left(\Phi-\Psi\right)+
\left(\bar{\nabla}_{k}\Phi\right)^{2}-\left(\bar{\nabla}_{k}\Psi\right)^
{2}+2\bar{\nabla}_{k}\Phi\bar{\nabla}^{k}\Psi\right]-\bar{\nabla}^{i}
\bar{\nabla}_{j}\left(\Phi-\Psi\right)\nn\\
&&-\bar{\nabla}^{i}\Phi\bar{\nabla}_{j}\Phi+\bar{\nabla}^{i}\Psi\bar{
\nabla}_{j}\Psi-\bar{\nabla}^{i}\Phi\bar{\nabla}_{j}\Psi-\bar{\nabla}_{j}
\Phi\bar{\nabla}^{i}\Psi=\bar{\nabla}^{i}\phi\bar{\nabla}_{j}\phi-\frac{
1}{3}\delta_{j}^{i}\left(\bar{\nabla}_{k}\phi\right)^{2}.\label{EE_ij_tf
} \eea At first order, we recover the familiar equality
$\Phi_{(1)}=\Psi_{(1)}$ that holds in the case of a perfect fluid,
regardless of the spatial curvature \cite{PeterUzan2009}. 
Eq.~(\ref{phi_1st_sol}) then simplifies to \begin{equation}\phi_{(1)}=\frac{2}{\bar{\phi}'}\lb\Phi_{(1)}'+\mathcal{H}\Phi_{(1)}
\rb.\label{phi_1st_sol_Phi_1st}
\end{equation}
We can now express $\phi'_{(1)}$ in
terms of $\Phi_{(1)}$ and its time derivatives by simply taking the
first time derivative of Equation (\ref{phi_1st_sol_Phi_1st}), \begin{equation}\phi_{(1)}'=\frac{2}{\bar{\phi}'}\left[\Phi_{(1)}''+\left(\mathcal{H}-
\frac{\bar{\phi}''}{\bar{\phi}'}\right)\Phi_{(1)}'+\left(\mathcal{H}'-
\mathcal{H}\frac{\bar{\phi}''}{\bar{\phi}'}\right)\Phi_{(1)}\right].
\label{phi_1st_prime_Phi_1st} \end{equation}
At second order, using $\Phi_{(1)}=\Psi_{(1)}$,
we obtain

\begin{equation}
\small{
\frac{1}{6}\delta_{j}^{i}\bar{\nabla}^{2}\left(\Phi_{(2)}-\Psi_{(2)}
\right)-\frac{1}{2}\bar{\nabla}^{i}\bar{\nabla}_{j}\left(\Phi_{(2)}-\Psi_
{(2)}\right)=2\left[\bar{\nabla}^{i}\Phi_{(1)}\bar{\nabla}_{j}\Phi_{(1)}
-\frac{1}{3}\delta_{j}^{i}\left(\bar{\nabla}_{k}\Phi_{(1)}\right)^{2}
\right]+\bar{\nabla}^{i}\phi_{(1)}\bar{\nabla}_{j}\phi_{(1)}-\frac{1}{3}
\delta_{j}^{i}\left(\bar{\nabla}_{k}\phi_{(1)}\right)^{2}}
\label{EE_ij_tf_2nd}
\end{equation}
By acting on the left- and right-hand sides of
(\ref{EE_ij_tf_2nd}) with $\bar{\nabla}_{i}\bar{\nabla}^{j}$ and using
$\left[\bar{\nabla}^{2},\bar{\nabla}_{i}\right]X\equiv
\bar{R}_{ij}\bar{\nabla}^{j}X$ with $\bar{R}_{ij}=2\Ka\gamma_{ij}$, we
obtain
\begin{equation}\Psi_{(2)}-\Phi_{(2)}=2F\left(\Psi_{(1)}\right)+F\left(\phi_{(1)}\right)
.\label{Phi_Psi_2nd_sol}
\end{equation}where
\begin{equation}
F\left(X\right)=\left(\bar{\nabla}^{2}\bar{\nabla}^{2}+3\Ka\bar{\nabla}^
{2}\right)^{-1}\left[3\left(\bar{\nabla}^{2}X\right)^{2}+\left(\bar{
\nabla}_{i}\bar{\nabla}_{j}X\right)^{2}+4\bar{\nabla}_{i}\bar{\nabla}^{2}
X\bar{\nabla}^{i}X+2\Ka\left(\bar{\nabla}_{i}X\right)^{2}\right].\label{
F_def_3}
\end{equation}
For later convenience, we also note that
\begin{equation}\bar{\nabla}^{2}F\left(X\right)=\left(\bar{\nabla}^{2}+3\Ka\right)^{-1}
\left[3\left(\bar{\nabla}^{2}X\right)^{2}+\left(\bar{\nabla}_{i}
\bar{\nabla}_{j}X\right)^{2}+4\bar{\nabla}_{i}\bar{\nabla}^{2}X\bar{\nabla}^{i
}X+2\Ka\left(\bar{\nabla}_{i}X\right)^{2}\right].\label{F_nabla2} \ee
\noindent The energy constraint equation $G_0^0=T_0^0$ reads \bea
&&-3\frac{\ex^{-2\Phi}}{a^{2}}\left(\mathcal{H}-\Psi'\right)^{2}-\frac{
\ex^{2\Psi}}{a^{2}}\left[2\bar{\nabla}^{2}\Psi-\left(\bar{\nabla}_{i}\Psi
\right)^{2}+3\Ka\right]=-\frac{\ex^{-2\Phi}}{2a^{2}}\left(\phi'\right)^{
2}-\frac{\ex^{2\Psi}}{2a^{2}}\left(\bar{\nabla}_{i}\phi\right)^{2}-V
\left(\phi\right),\label{EE_00} \eea and yields the first Friedman
equation at zeroth order, \begin{equation}\mathcal{H}^{2}=\frac{1}{3}\lb\frac{\bar{\phi}'^{2}}{2}+a^{2}\bar{V}\rb-
\Ka.\label{EE_00_bg} \ee
The trace equation $\mrm{tr}\lb G_j^i\rb=\mrm{tr}\lb T_j^i\rb$ reads
{\small
\bea
&&3\frac{\ex^{-2\Phi}}{a^{2}}\left[-2\mathcal{H}'-\left(\mathcal{H}-2\Phi'
-3\Psi'\right)\left(\mathcal{H}-\Psi'\right)+2\Psi''\right]+\frac{\ex^{2
\Psi}}{a^{2}}\left[2\bar{\nabla}^{2}\left(\Phi-\Psi\right)+2\left(\bar{
\nabla}_{i}\Phi\right)^{2}+\left(\bar{\nabla}_{i}\Psi\right)^{2}-2\bar{
\nabla}_{i}\Phi\bar{\nabla}^{i}\Psi-3\Ka\right]\nn\\ &&
=\frac{3}{2}\frac{\ex^{-2\Phi}}{a^{2}}\left(\phi'\right)^{2}-\frac{\ex^{
2\Psi}}{2a^{2}}\left(\bar{\nabla}_{k}\phi\right)-3V\left(\phi\right),
\label{EE_ij_tr} \eea }
and its zeroth order can be combined with the
zeroth order energy constraint equation through 
\begin{equation}
G_{0}^{0}-\frac{1}{3}\mathrm{tr}\,G_{j}^{i}=T_{0}^{0}-\frac{1}{3}\mathrm
{tr}\,T_{j}^{i} 
\end{equation}
to yield the second Friedman equation
\begin{equation}\mathcal{H}'=-\frac{1}{3}\lb
\bar{\phi}'^{2}-a^{2}\bar{V}\rb.
\label{fried2} 
\end{equation}
Combining
the zeroth order trace equation with the zeroth order energy constraint
equation as 
\begin{equation}
G_{0}^{0}-\mathrm{tr}\,G_{j}^{i}=T_{0}^{0}-\mathrm{tr}\,T_{j}^{i},
\ee
we obtain
\begin{equation}3\mcl H'+\phi'^2-a^2\bar{V}=0,
\label{EE_00_ij_tr_comb3}
\end{equation}
which can be combined with the time derivative of (\ref{EE_00_bg})
to yield the Klein-Gordon equation
\begin{equation}
\bar{\phi}''+2\mathcal{H}\bar{\phi}'+a^{2}\bar{V}_{,\phi}=0.
\label{sf_eom_bg} 
\end{equation}
The time derivative of the latter equation then yields
\begin{equation}
a^{2}\bar{V}_{,\phi\phi}=2\left(2\mathcal{H}^{2}-\mathcal{H}'\right)-
\frac{\bar{\phi}'''}{\bar{\phi}'}\,,\label{V_der2_sol}
\end{equation}
which can be
used to replace the third time derivative of $\bar\phi$ by the second
$\phi$-derivative of the potential $V(\phi)$.
Using the following combination of the energy constraint and
trace equations
\begin{equation}
G_{0}^{0}+\frac{1}{3}\mathrm{tr}\,G_{j}^{i}=T_{0}^{0}+\frac{1}{3}\mathrm
{tr}\,T_{j}^{i} 
\end{equation}
at first order together with
(\ref{phi_1st_sol_Phi_1st}), (\ref{sf_eom_bg}) and
$\Phi_{(1)}=\Psi_{(1)}$ yields the dynamical equation for the first
order scalar metric perturbation
\begin{equation}
\Psi_{(1)}''+2\left(\mathcal{H}-\frac{\bar{\phi}''}{\bar{\phi}'}\right)
\Psi_{(1)}'-\bar{\nabla}^{2}\Psi_{(1)}+2\left(\mathcal{H}'-\mathcal{H}
\frac{\bar{\phi}''}{\bar{\phi}'}-2\Ka\right)\Psi_{(1)}=0\,.
\label{Psi_1st_dyn}
\end{equation}
At second order, the above combination of the
energy contraint and trace equations yields 
\bea 
&& 2\left(2\mathcal{H}^{2}+\mathcal{H}'\right){\Phi_{(2)}}+\mathcal{H}
{\Phi_{(2)}'}-4\Ka{\Psi_{(2)}}+5\mathcal{H}{\Psi_{(2)}'}+{\Psi_{(2)}''}-
\nn \\ &&
4\left(2\Ka+2\mathcal{H}^{2}+\mathcal{H}'\right)\Psi_{(1)}^{2}-8\Psi_{(1
)}'^{2}+\Psi_{(1)}\left(-24\mathcal{H}\Psi_{(1)}'-4{\Psi_{(1)}''}-4\bar{
\nabla}^{2}\Psi_{(1)}\right)\nn+ \\ &&
\frac{1}{3}\left[3a^{2}\bar{V}_{,\phi\phi}\phi_{(1)}^{2}+3a^{2}\bar{V}_{
,\phi}{\phi_{(2)}}+{\bar{\nabla}^{2}\Phi_{(2)}}-4{\bar{\nabla}^{2}\Psi_{
(2)}}+2\left(\bar{\nabla}_{i}\phi_{(1)}\right)^{2}+4\left(\bar{\nabla}_{
i}\Psi_{(1)}\right)^{2}\right]=0\label{EE_dyn_2nd_ori} \eea Using this
expression, we may now derive a dynamical equation for $\Psi_{(2)}$ with
a source term involving quadratic combinations of $\Psi_{(1)}$, its first time
derivative, and its spatial derivatives.  The second time derivative of
$\Psi_{(1)}$ can be eliminated using (\ref{Psi_1st_dyn}).  The first
order scalar field perturbation $\phi_{(1)}$ can be eliminated using
(\ref{phi_1st_sol_Phi_1st}).  The second order perturbation $\Phi_{(2)}$
is related to $\Psi_{(2)}$ through (\ref{Phi_Psi_2nd_sol}) with which we
also obtain 
\begin{equation}\Phi_{(2)}'={\Psi_{(2)}'}-\partial_{\eta}\left[2F\left(\Psi_{(1)}\right)
+F\left(\phi_{(1)}\right)\right],
\label{Phi_2nd_prime_sim}
\end{equation}
where \begin{equation}\partial_{\eta}\left[F\left(X\right)\right]=
2\left(\bar{\nabla}^{2}\bar{\nabla}^{2}+3\Ka\bar{\nabla}^{2}\right)^{-1}
\bar{\nabla}_{i}\bar{\nabla}^{j}\left(3\bar{\nabla}^{i}X'\bar{
\nabla}_{j}X-\delta_{j}^{i}\bar{\nabla}_{k}X'\bar{\nabla}^{k}X
\right).\label{F_der}
\end{equation}
Finally, using (\ref{phi_2nd_sol_ori}) and
(\ref{Phi_Psi_2nd_sol}), we obtain \begin{equation}\phi_{(2)}=\frac{2}{\bar{\phi}'}\left\{{\Psi_{(2)}'}+\mathcal{H}{\Psi_{(
2)}}-\mathcal{H}\left[2F\left(\Psi_{(1)}\right)+F\left(\phi_{(1)}\right)
\right]-\bar{\nabla}^{-2}\bar{\nabla}^{i}\left(2\Psi_{(1)}'\bar{\nabla}_
{i}\Psi_{(1)}+\phi_{(1)}'\bar{\nabla}_{i}\phi_{(1)}\right)\right\},
\label{phi_2nd_sim} \end{equation}from which $\phi_{(1)}$ and $\phi_{(1)}'$ can be
eliminated using (\ref{phi_1st_sol_Phi_1st}) and
(\ref{phi_1st_prime_Phi_1st}), respectively.  Using these substitutions and the background equations, we obtain (\ref{Psi_2_eom}) with source term (\ref{source_2nd}).
 
 \section{Correlation function and power spectrum}
\label{powerspc}
 
 \noindent  Assuming statistical homogeneity and isotropy, the
 correlation function $\xi(\alpha)$ of two fluctuations $\delta(\Omega_1)$ and
 $\delta(\Omega_2)$ at points of angular coordinates $\Omega_1$ and
 $\Omega_2$ respectively on the 3-sphere which are separated by an angle
 $\alpha$ can be decomposed in terms of hyperspherical harmonics $Q_{n\ell m}$ as
\begin{equation}
\la
\xi(\alpha)=\la\delta(\Omega_1)\delta(\Omega_2)\ra=\sum_n\frac{n+1}{2\pi^2
 }C_n^1(\cos\alpha)\xi_n=\sum_n\xi_n\sum_{\ell m}Q^*_{n\ell
 m}(\Omega_1)Q_{n\ell m}(\Omega_2)\,, \label{eq:correlation1}
\end{equation}
where in the intermediate step $C_n^{(1)}$ is the Gegenbauer polynomial of degree $n$ and order $1$ and where we
 have used the addition theorem for the functions $Q_{n\ell m}$ (see
 Appendix \ref{FormFactor}).  $\xi(\alpha)$ may also be written as
\begin{equation}
\la
 \delta(\Omega_1)\delta(\Omega_2)\ra=\xi(\alpha)=\sum_{n_1\ell_1
 m_1}\sum_{n_2\ell_2 m_2}Q_{n_1\ell_1 m_1}^*Q_{n_2\ell_2 m_2}\la
 \delta_{n_1\ell_1m_1} \delta_{n_1\ell_1m_1}\ra\,.
\label{eq:correlation2}
 \end{equation}
By equating (\ref{eq:correlation1}) and (\ref{eq:correlation2}) and
 integrating with respect to $\dd \Omega_1$ and $\dd\Omega_2$ (with
 $\dd\Omega_i=\sin\theta_i\sin^2\chi_i\,\dd\phi_i\,\dd\theta_i\,\dd
\chi_i$), we obtain, using (\ref{Q_nlm_orthonormal})
\begin{equation}
\la
 \delta_{n\ell m}^*\delta_{n'\ell'm'}\ra=\xi_n\delta_{nn'}\delta_{\ell
 \ell'}\delta_{mm'}. \label{eq:2ptfunc}
\end{equation}
We may follow \cite{Lyth:1990dh} in defining the power spectrum as
\begin{equation}
\mcl P_\delta(q)=\frac{q(q^2-1)}{2\pi^2}\xi_n,
\label{eq:Pq}
\end{equation}
where $q=n+1$, so that the autocorrelation function reads
\begin{equation}\la
 \delta^2\ra=\sum_n\frac{q}{q^2-1}\mcl P_{\delta}(q),
\end{equation}
in analogy with the flat space definition where $\delta=\int \mcl P_{\delta}(k)\dd
 k/k$.

\section{Mode functions in $\Ka>0$ spacetime}
\label{FormFactor}

As usual, it is convenient to work in $k$-space instead of configuration
space. In our case, since the spatial volume has constant positive
curvature, we shall make use of the 3-dimensional hyperspherical
harmonics $Q_{n\ell m}(\chi,\theta,\phi)$, which are given by
(\ref{Q_nlm_def})-(\ref{Pi_nl_def}). Any real scalar field
$f(\bm{x})\equiv f(\chi,\theta,\phi)$ can then be expanded in terms of
$Q_{n\ell m}(\chi,\theta,\phi)$, using the normalization given in
(\ref{Q_nlm_orthonormal}), as \begin{equation}
f\left(\bm{x}\right)=\sum_{n,l,m}f_{n\ell m}Q_{n\ell
m}\left(\bm{x}\right),\qquad\text{with}\qquad f_{n\ell m}=\int
\dd^{3}x\sqrt{h}Q_{n\ell
m}^{\ast}\left(\bm{x}\right)f\left(\bm{x}\right), \end{equation} where
$\sqrt{h}=f_\Ka^2 (\chi) \sin\theta$, with $f_\Ka \equiv \Ka^{-1/2}
\sin\left(\sqrt{\Ka} \chi\right)$.  The $Q_{n\ell m}$'s are the
eigenmodes of the Laplacian $\Delta \equiv \nabla^2$: \begin{equation}
\Delta Q_{n\ell m} = -k^2Q_{n\ell m}=-\Ka n(n+2) Q_{n\ell m},\qquad
n\geq 1. \end{equation} 
As done in Appendix \ref{powerspc}, it can be convenient to introduce $q=n+1$, with
$q$ running from $0$ to $\infty$ so that \begin{equation}\Delta Q_{n\ell m}=-\Ka
(q^2-1)Q_{n\ell m}. \end{equation}\noindent The harmonics in $\mathbb{S}^3$ are
given by hyperspherical harmonics which can be expressed as
\begin{equation}
Q_{n\ell m}\left(\chi,\theta,\phi\right)=R_{nl}\left(\chi\right)Y_{\ell
m}\left(\theta,\phi\right),
\label{Q_nlm_def}
\end{equation}
where the $Y_{\ell m}(\theta,\phi)$'s are the standard spherical harmonics in
$\mathbb{S}^2$ and
\begin{equation}
R_{n\ell}\left(\chi\right) =
\sqrt{\frac{\left(n+1\right)\left(n+\ell+1\right)!}{\left(n-\ell\right)!
}}\sqrt{\frac{\Ka}{f_\Ka
\left(\chi\right)}}P_{n+\frac{1}{2}}^{-\ell-\frac{1}{2}}\left[\cos\left(
\sqrt{\Ka}\chi\right)\right].
\label{Pi_nl_def}
\end{equation}
The associated Legendre
functions of half-integer degree and order read \begin{equation}P_{n+1/2}^{-\ell-1/2}(x)=(-1)^{n-\ell}\frac{(1-x^2)^{-\ell/2-1/4}}{2^{n+
\ell/2}\Gamma(n+3/2)}\frac{\dd^{n-\ell}}{\dd x^{n-\ell}}(1-x^2)^{n+1/2}\,,
\end{equation}and are related to the Gegenbauer polynomials according to the
relation \begin{equation}C_{n-\ell}^{\ell+1}(x)=2^{-\ell}\lb\frac{\pi}{2}\rb^{1/2}(1-x^2)^{-\ell/
2-1/4}\frac{\Gamma(n+\ell+2)}{(n-\ell)!\Gamma(\ell+1)}P_{n+1/2}^{-\ell-1
/2}(x). \end{equation}With the choice of normalisation made for
$\Pi_{n\ell}(\chi)$, the quantity $Q_{n\ell
m}\left(\chi,\theta,\phi\right)$ satisfies the orthogonality condition
\begin{equation}{\label{Q_nlm_orthonormal}} \int
\dd^{3}x\sqrt{h}Q_{n\ell
m}^{\ast}\left(\bm{x}\right)Q_{n'l'm'}\left(\bm{x}\right)=\delta_{nn'}
\delta_{ll'}\delta_{mm'}. \end{equation} The addition theorem for
hyperspherical harmonics on the 3-sphere reads \begin{equation}\sum_{\ell
m}Q^*_{n\ell m}(\Omega_1)Q_{n\ell
m}(\Omega_2)=\frac{n+1}{2\pi^2}C_n^1(\cos\alpha). \end{equation}Here,
$\Omega_i\equiv (\chi_i,\theta_i,\phi_i)$ and $\alpha$ is the angle
between the two direction defined by the angles $\Omega_1$ and
$\Omega_2$; $C_n^1(\cos \alpha)$ is the Gegenbauer polynomial of degree
$n$ and order $1$.

 Since the $Q_{n\ell m}$'s form a complete and orthogonal basis in
 $\mathbb{S}^3$, the product of any two $Q_{n\ell m}$'s can be expanded
 as a summation over $Q_{n\ell m}$: \begin{equation}
 Q_{\bm{p}_{1}}\left(\bm{x}\right)Q_{\bm{p}_{2}}\left(\bm{x}\right)=
\sum_{\bm{k}}\mathcal{G}_{\bm{k},\bm{p}_{1},\bm{p}_{2}}Q_{\bm{k}}\left(
\bm{x}\right), \end{equation} where, using (\ref{Q_nlm_orthonormal}),
 \begin{equation}{\label{Gaunt_coeff_S3}}
 \mathcal{G}_{\bm{k},\bm{p}_{1},\bm{p}_{2}}\equiv\int
 \dd^{3}x\sqrt{h}Q_{\bm{k}}^{\ast}\left(\bm{x}\right)Q_{\bm{p}_{1}}\left
 (\bm{x}\right)Q_{\bm{p}_{2}}\left(\bm{x}\right). \end{equation} These
 expansion coefficients are the generalization of the Gaunt's
 coefficients of spherical harmonics in $\mathbb{S}^2$ to
 $\mathbb{S}^3$, which are purely geometric quantities in
 $\mathbb{S}^3$.  The explicit form of this integral is given
 by~\cite{Wen:1985mp} \bea \mcl G_{\bm k_1\bm k_2\bm k_3}= I_1 \lb
 \frac{1}{2}\,;\, \begin{array}{ccc} \ell_1&\ell_2&\ell_3\\ m_1&m_2&m_3
 \end{array} \rb \cdot I_2 \lb 1\,;\, \begin{array}{ccc} n_1&n_2&n_3\\
 \ell_1&\ell_2&\ell_3 \end{array} \rb . \eea In this expression, the
 quantity $I_1$ is given by \bea I_1 \lb \frac{1}{2}\,;\,
 \begin{array}{ccc} \ell_1&\ell_2&\ell_3\\ m_1&m_2&m_3 \end{array} \rb
 =\lsb\frac{(2\ell_1+1)(2\ell_2+1)(2\ell_3+1)}{4\pi}\rsb^{1/2} \lb
 \begin{array}{ccc} \ell_1&\ell_2&\ell_3\\ 0&0&0 \end{array} \rb \cdot
 \lb \begin{array}{ccc} \ell_1&\ell_2&\ell_3\\ m_1&m_2&m_3 \end{array}
 \rb\nn, \eea where the Wigner 3-j symbols are non-zero provided \begin{equation} \ell_1+\ell_2+\ell_3\,\,\text{is even},\quad m_1+m_2+m_3=0,\quad
 \left|m_i\right|\leq \ell_i,\quad\text{and}\quad
 \left|\ell_i-\ell_j\right|\leq \ell_k\leq \ell_i+\ell_j, \end{equation}while the
 quantity $I_2$ reads \bea I_2 \lb 1\,;\, \begin{array}{ccc}
 n_1&n_2&n_3\\ \ell_1&\ell_2&\ell_3 \end{array} \rb &=&
 \left[\prod_{i=1}^3F_0(n_i,\ell_i)\right]\sum_{t_i=0}^{(n_i-\ell_i)/2}
\left[\prod_{i=1}^{3}F_1(n_i,\ell_i,t_i) \right]\nn\\
&&\sum_{\tau_1\tau_2}(2\tau_1+1)(2\tau_2+1) \lb \begin{array}{ccc}
 \nu_1&\nu_2&\tau_1\\ 0&0&0 \end{array} \rb^2\cdot \lb
 \begin{array}{ccc} \nu_3&\tau_2&\tau_1\\ 0&0&0 \end{array} \rb^2\cdot
 F_2(M,\tau_2),\nn \eea where $\tau_1$ and $\tau_2$ take on all allowed
 values of the Wigner 3-j symbols.  In these expressions, \begin{equation} M=\ell_1+\ell_2+\ell_3,\qquad
 \nu_i=n_i-\ell_i-2t_i,\quad\text{and}\quad 0\leq t_i\leq
 \frac{n_i-\ell_i}{2}, \end{equation}and $F_0$, $F_1$ and $F_2$ are given by \bea
 F_0(n_i,\ell_i)&=&\lsb\frac{2^{2\ell_i}\Gamma^2\lb \ell_i+1\rb
 \Gamma\lb n_i-\ell_i+1\rb\lb 2n_i+2\rb}{\pi\Gamma\lb
 n_i+\ell_i+2\rb}\rsb^{1/2}\nn,\\
F_1(n_i,\ell_i,t_i)&=&\frac{\sqrt{\pi}\lb
 n_i+\ell_i-2t_i+\frac{1}{2}\rb\Gamma\lb
 \ell_i+t_i+\frac{1}{2}\rb\Gamma\lb n_i-t_i+1\rb}{\Gamma\lb
 1+t_i\rb\Gamma\lb \ell_i+\frac{1}{2}\rb\Gamma\lb
 n_i-\ell_i-t_i+\frac{3}{2}\rb\Gamma\lb \ell_i+1\rb}\nn,\\
F_2(M,\tau_2)&=&\frac{\pi\Gamma^2\lb
 \frac{M}{2}+\frac{3}{2}\rb}{\Gamma\lb
 \frac{M}{2}+\frac{\tau_2}{2}+2\rb\Gamma\lb\frac{M}{2}-\frac{\tau_2}{2}+
 \frac{3}{2}\rb\Gamma\lb\frac{\tau_2}{2}+1\rb\Gamma\lb\frac{1}{2}-\frac{
 \tau_2}{2}\rb}. \eea The quantity $I_2$ is non-zero provided the
 relations \begin{equation}n_1+n_2+n_3\quad\text{is
 even,}\quad\ell_i<n_i,\quad\text{and}\quad\left|n_i-n_j\right|\leq
 n_k\leq n_i+n_j \end{equation}hold.

\section{Source terms in hyperspherical harmonic space}
\label{sourcetermkspace}

In hyperspherical harmonic space, the right-hand side of the dynamical
equation (\ref{Psi_2_eom}), \ie $\mcl S_{(2)}$ in (\ref{source_2nd}),
will be the sum of ten terms each expressible in the generic form
\begin{equation}
\sum_{\bm{p_1} \bm{p_2}}\mcl
F(\eta)\,\mathcal{G}_{\bm{k},\bm{p_1},\bm{p_2}}\,
K(k,p_1,p_2)\,\frac{\partial^i\Psi_{\bm{p_1}}}{\partial\eta^i}
\frac{\partial^j \Psi_{\bm{p_2}}}{\partial \eta^j}\qquad\text{with}\qquad i,j=0
\text{ or }1,
\end{equation}
where  $\mcl F(\eta)$ is a function of the background
only, $\mathcal{G}_{\bm{k},\bm{p_1},\bm{p_2}}$ is the geometrical form
factor of Appendix \ref{FormFactor}, and $K(k,p_1,p_2)$ is a momentum
factor coming from the spatial derivatives acting on $\Psi_1$, with
$k^2=n(n+2)$ and $p_i^2=n_i(n_i+2)$.

Making use of the following set of harmonic transformations (HT's) on
$\mathbb{S}^{3}$, \bea fg\xrightarrow{\mrm{HT}}\left[fg\right]_{\bm{k}}
& = &
\sum_{\bm{p}_{1},\bm{p}_{2}}\mathcal{G}_{\bm{k},\bm{p}_{1},\bm{p}_{2}}
f_{\bm k_1}g_{\bm k_2},\label{quadratic_trans}\\
\nabla_{i}f\nabla^{i}g\xrightarrow{\mrm{HT}}\left[\nabla_{i}f\nabla^{i}
g\right]_{\bm k} & = &
\sum_{\bm{p}_{1},\bm{p}_{2}}\mathcal{G}_{\bm{k},\bm{p}_{1},\bm{p}_{2}}
\frac{1}{2}\left(p_{1}^{2}+p_{2}^{2}-k^{2}\right)f_{\bm k_1}g_{\bm
k_2},\label{quadratic_trans_deri_1}\\
\nabla^{2}\left(fg\right)\xrightarrow{\mrm{HT}}\left[\nabla^{2}\left(fg
\right)\right]_{\bm k} & = &
\sum_{\bm{p}_{1},\bm{p}_{2}}\mathcal{G}_{\bm{k},\bm{p}_{1},\bm{p}_{2}}
\left(-k^{2}\right)f_{\bm k_1}g_{\bm
k_2},\label{quadratic_trans_deri_2}\\
\nabla_{i}\left(f\nabla^{i}g\right)\xrightarrow{\mrm{HT}}\left[\nabla_{
i}\left(f\nabla^{i}g\right)\right]_{\bm k} & = &
\sum_{\bm{p}_{1},\bm{p}_{2}}\mathcal{G}_{\bm{k},\bm{p}_{1},\bm{p}_{2}}
\frac{1}{2}\left(p_{1}^{2}-p_{2}^{2}-k^{2}\right)f_{\bm k_1}g_{\bm
k_2},\label{quadratic_trans_deri_3}\\
\nabla_{i}\nabla_{j}f\nabla^{i}\nabla^{j}g\xrightarrow{\mrm{HT}}\left[
\nabla_{i}\nabla_{j}f\nabla^{i}\nabla^{j}g\right]_{\bm k} & = &
\sum_{\bm{p}_{1},\bm{p}_{2}}\mathcal{G}_{\bm{k},\bm{p}_{1},\bm{p}_{2}}
\frac{1}{4}\left(p_{1}^{2}+p_{2}^{2}-k^{2}\right)^{2}f_{\bm k_1}g_{\bm
k_2}, \eea where, in the above, we use $\bm{k}$ to represent
$\left\{n,\ell,m\right\}$ and $\bm{p}_{i}$ to represent $\left\{
n_{i},\ell_{i},m_{i}\right\}$, and introducing
\bea
P_+=\frac{p_1^2+p_2^2}{k^2},\qquad
P_-=\frac{p_1^2-p_2^2}{k^2},\quad
\Xi=\frac{k^2\lb 2P_+-3P_-^2\rb+
\Ka\lb 4P_+-1\rb}{k^2-3\Ka}\nn,
\label{Ppm_Pi_def}
\eea
the ten terms in $\mcl S_{(2)}$
transform as {\scpt \bea \mrm{[1]}&\quad&
4\left(2\mathcal{H}^{2}-\mathcal{H}'+2\mathcal{H}\frac{\bar{\phi}''}{
\bar{\phi}'}+6\Ka\right)\Psi_{(1)}^{2}\xrightarrow{\mrm{HT}}4\left(2
\mathcal{H}^{2}-\mathcal{H}'+2\mathcal{H}\frac{\bar{\phi}''}{\bar{\phi}'}
+6\Ka\right){\Psi_{(1)}\left(\bm{p}_{1},\eta\right)\Psi_{(1)}\left(\bm
{p}_{2},\eta\right)},\\ \mrm{[2]}&\quad&
8\Psi_{(1)}'^{2}\xrightarrow{\mrm{HT}}8{\Psi_{(1)}'\left(\bm{p}_{1},
\eta\right)\Psi_{(1)}'\left(\bm{p}_{2},\eta\right)},\\ \mrm{[3]}&\quad&
8\left(2\mathcal{H}+\frac{\bar{\phi}''}{\bar{\phi}'}\right)\Psi_{(1)}
\Psi_{(1)}'\xrightarrow{\mrm{HT}}8\left(2\mathcal{H}+\frac{\bar{\phi}''}
{\bar{\phi}'}\right){\Psi_{(1)}\left(\bm{p}_{1},\eta\right)\Psi_{(1)}'
\left(\bm{p}_{2},\eta\right)},\\ \mrm{[4]}&\quad&
8\Psi_{(1)}\bar{\nabla}^{2}\Psi_{(1)}\xrightarrow{\mrm{HT}}-4k^{2}P_{+}
{\Psi_{(1)}\left(\bm{p}_{1},\eta\right)\Psi_{(1)}\left(\bm{p}_{2},\eta
\right)},\\ \mrm{[5]}&\quad&
-\frac{4}{3}\left(\bar{\nabla}_{i}\Psi_{(1)}\right)^{2}\xrightarrow{
\mrm{HT}}\frac{2}{3}k^{2}\left(1-P_{+}\right){\Psi_{(1)}\left(\bm{p}_{1}
,\eta\right)\Psi_{(1)}\left(\bm{p}_{2},\eta\right)},\\ \mrm{[6]}&\quad&
-\left[2\left(2\mathcal{H}^{2}-\mathcal{H}'\right)-\frac{\bar{\phi}'''}
{\bar{\phi}'}\right]\phi_{(1)}^{2}\xrightarrow{\mrm{HT}}-\lsb
2\left(2\mathcal{H}^{2}-\mathcal{H}'\right)-\frac{\bar{\phi}'''}{\bar{
\phi}'}\right]\frac{4}{\left(\bar{\phi}'\right)^{2}}\Big[\Psi_{(1)}'
\left(\bm{p}_{1},\eta\right)\Psi_{(1)}'\left(\bm{p}_{2},\eta\right)+2
\mathcal{H}\Psi_{(1)}\left(\bm{p}_{1},\eta\right)\Psi_{(1)}'\left(\bm{p}
_{2},\eta\right)\nn\\
&&+\mathcal{H}^{2}{\Psi_{(1)}\left(\bm{p}_{1},\eta\right)\Psi_{(1)}
\left(\bm{p}_{2},\eta\right)}\Big],\\ \mrm{[7]}&\quad&
-\frac{2}{3}\left(\bar{\nabla}_{i}\phi_{(1)}\right)^{2}\xrightarrow{
\mrm{HT}}\frac{4}{3\left(\bar{\phi}'\right)^{2}}k^{2}\left(1-P_{+}\right
)\Big[{\Psi_{(1)}'\left(\bm{p}_{1},\eta\right)\Psi_{(1)}'\left(\bm{p}_{
2},\eta\right)}+2\mathcal{H}{\Psi_{(1)}\left(\bm{p}_{1},\eta\right)
\Psi_{(1)}'\left(\bm{p}_{2},\eta\right)}\nn\\
&&+\mathcal{H}^{2}{\Psi_{(1)}\left(\bm{p}_{1},\eta\right)\Psi_{(1)}
\left(\bm{p}_{2},\eta\right)}\Big],\\ \mrm{[8]}&\quad&
-2\left(\frac{\bar{\phi}''}{\bar{\phi}'}+2\mathcal{H}\right){\bar{
\nabla}^{-2}}{\bar{\nabla}^{i}\left(2\Psi_{(1)}'\bar{\nabla}_{i}\Psi_{(1
)}+\phi_{(1)}'\bar{\nabla}_{i}\phi_{(1)}\right)}\xrightarrow{\mrm{HT}}
\left(\frac{\bar{\phi}''}{\bar{\phi}'}+2\mathcal{H}\right)\Bigg\{\frac{
4}{\lb\bar{\phi}'\right)^{2}}\left(\mathcal{H}-\frac{\bar{\phi}''}{\bar
{\phi}'}\rb\Psi_{(1)}'\left(\bm{p}_{1},\eta\right)\Psi_{(1)}'\left(\bm{
p}_{2},\eta\rb\nn\\
&&+\left\{-\left(1+P_{-}\right)\left[2-\frac{4\mathcal{H}}{\left(\bar{
\phi}'\right)^{2}}\left(\mathcal{H}-\frac{\bar{\phi}''}{\bar{\phi}'}
\right)\right]+\frac{4}{\left(\bar{\phi}'\right)^{2}}\left[\mathcal{H}'-
\mathcal{H}\frac{\bar{\phi}''}{\bar{\phi}'}-4\Ka+\frac{k^{2}}{2}\left(
P_{+}+P_{-}\right)\right]\left(1-P_{-}\right)\right\}{\Psi_{(1)}\left(
\bm{p}_{1},\eta\right)\Psi_{(1)}'\left(\bm{p}_{2},\eta\right)}\nn\\
&&+\frac{4\mathcal{H}}{\left(\bar{\phi}'\right)^{2}}\lsb\mathcal{H}'-
\mathcal{H}\frac{\bar{\phi}''}{\bar{\phi}'}-4\Ka+\frac{k^{2}}{2}\left(P_
{+}-P_{-}^{2}\right)\rsb{\Psi_{(1)}\left(\bm{p}_{1},\eta\right)\Psi_{(1
)}\left(\bm{p}_{2},\eta\right)}\Bigg\},\\ \mrm{[9]}&\quad&
\left[2\left(\mathcal{H}'-\mathcal{H}\frac{\bar{\phi}''}{\bar{\phi}'}
\right)+\frac{1}{3}\bar{\nabla}^{2}\right]\left[2F\left(\Psi_{(1)}\right
)+F\left(\phi_{(1)}\right)\right]\xrightarrow{\mrm{HT}}\left[2\left(
\mathcal{H}'-\mathcal{H}\frac{\bar{\phi}''}{\bar{\phi}'}\right)-\frac{1}
{3}k^{2}\right]\left(1+\Xi\right)\nn\\
&&\times\left[\left(\frac{1}{2}+\frac{\mathcal{H}^{2}}{\left(\bar{\phi}
'\right)^{2}}\right){\Psi_{(1)}\left(\bm{p}_{1},\eta\right)\Psi_{(1)}
\left(\bm{p}_{2},\eta\right)}+\frac{1}{\left(\bar{\phi}'\right)^{2}}{
\Psi_{(1)}'\left(\bm{p}_{1},\eta\right)\Psi_{(1)}'\left(\bm{p}_{2},\eta
\right)}+\frac{2\mathcal{H}}{\left(\bar{\phi}'\right)^{2}}{\Psi_{(1)}
\left(\bm{p}_{1},\eta\right)\Psi_{(1)}'\left(\bm{p}_{2},\eta\right)}
\right],\\
\mrm{[10]}&\quad&\mathcal{H}\left[2F\left(\Psi_{(1)}\right)+F\left(
\phi_{(1)}\right)\right]'\xrightarrow{\mrm{HT}}\mathcal{H}\left(1+\Xi
\right)\Bigg\{\left[1-\frac{2}{\left(\bar{\phi}'\right)^{2}}\left(
\mathcal{H}^{2}-2\mathcal{H}\frac{\bar{\phi}''}{\bar{\phi}'}+\mathcal{H}
'-4\Ka+\frac{k^{2}}{2}\left[P_{+}+P_{-}\right]\right)\right]{\Psi_{(1)}
\left(\bm{p}_{1},\eta\right)\Psi_{(1)}'\left(\bm{p}_{2},\eta\right)}\nn\\
&&-\frac{2}{\left(\bar{\phi}'\right)^{2}}\left(\mathcal{H}-\frac{\bar{
\phi}''}{\bar{\phi}'}\right){\Psi_{(1)}'\left(\bm{p}_{1},\eta\right)
\Psi_{(1)}'\left(\bm{p}_{2},\eta\right)}-\frac{2\mathcal{H}}{\left(\bar{
\phi}'\right)^{2}}\left(\mathcal{H}'-\mathcal{H}\frac{\bar{\phi}''}{
\bar{\phi}'}-4\Ka+\frac{1}{2}k^{2}P_{+}\right){\Psi_{(1)}\left(\bm{p}_{1},
\eta\right)\Psi_{(1)}\left(\bm{p}_{2},\eta\right)}\Bigg\}.
\eea
}
Collecting the coefficients multiplying the terms of type
$\Psi_1\Psi_2$, $\Psi_1\Psi_2'$, and $\Psi_1'\Psi_2'$, we obtain a set
of three coefficients $\mcl C_i$ (with $i=1\,\mrm{to}\,3$) which we
decompose as
\begin{equation}\mcl C_i=\mcl C_{i1}(\eta)+\mcl
C_{i2}(\Xi,P_+,P_-,\eta)\nn,
\end{equation}
where
{\scpt \bea \frac{\mcl
C_{11}}{\mcl H^2}&=&\lb 8-3\frac{\mcl H'}{\mcl H^2}+7\frac{\phi''}{\mcl
H\phi'}\rb+\frac{4\mcl H^2}{\phi'^2}\lsb 4\lb \frac{\mcl H'}{\mcl
H^2}-1\rb+\frac{\phi^{(3)}}{\mcl H^2\phi'}+\frac{\phi''}{\mcl H\phi'}\lb
\frac{\mcl H'}{\mcl H^2}-2-\frac{\phi''}{\mcl
H\phi'}\rb\rsb+\frac{8\Ka}{\mcl H^2}\lsb 3-\frac{\mcl H^2}{\phi'^2}\lb
3+2\frac{\phi''}{\mcl H\phi'}\rb\rsb,\nn\\ \frac{\mcl C_{12}}{\mcl
H^2}&=&\frac{k^2}{\mcl H^2}\lcb\frac{1}{2}+\frac{\mcl H^2}{\phi'^2}\lsb
1-\lb 4+\frac{2\phi''}{\mcl H\phi'}\rb P_-^2+\lb
\frac{5}{3}-\frac{14}{3}\frac{\phi'^2}{\mcl H^2}+2\frac{\phi''}{\mcl
H\phi'}\rb P_+-\frac{1}{3}\lb 1+\frac{1}{2}\frac{\phi'^2}{\mcl
H^2}\rb\Xi\rsb\rcb+\lb \frac{\mcl H'}{\mcl H^2}-\frac{\phi''}{\mcl
H\phi'}+8\frac{\mcl H^2}{\phi'^2}\frac{\Ka}{\mcl H^2}\rb\Xi,\nn\\
\frac{\mcl C_{21}}{\mcl H}&=&\lb 13+6\frac{\phi''}{\mcl
H\phi'}\rb+\frac{\mcl H^2}{\phi'^2}\lsb 8\frac{\phi^{(3)}}{\mcl
H^2\phi'}+26\lb \frac{\mcl H'}{\mcl H^2}-1\rb+\frac{2\phi''}{2\mcl
H\phi'}\lb\frac{\mcl H'}{\mcl H^2}-6-4\frac{\phi''}{\mcl
H\phi'}\rb-4\frac{\Ka}{\mcl H^2}\lb 6+4\frac{\phi''}{\mcl
H\phi'}\rb\rsb,\nn\\ \frac{\mcl C_{22}}{\mcl H}&=&\frac{4}{3}\frac{\mcl
H^2}{\phi'^2}\frac{k^2}{\mcl H^2}\lb 1-P_+\rb+2\frac{\mcl
H^2}{\phi'^2}\lb 2+\frac{\phi''}{\mcl H\phi'}\rb\lsb\lb 1-\frac{\mcl
H'}{\mcl H^2}-\frac{\phi'^2}{2\mcl H^2}+\frac{4\Ka}{\mcl H^2}\rb
P_-+\frac{k^2}{2\mcl H^2}\lb P_++P_-\rb\lb 1-P_-\rb\rsb+2\Xi\frac{\mcl
H^2}{\phi'^2}\lb \frac{\mcl H'}{\mcl H^2}-\frac{\phi''}{\mcl
H\phi'}-\frac{k^2}{6\mcl H^2}\rb\nn\\ &&-\frac{\mcl
H^2}{\phi'^2}\frac{2k^2}{6\mcl H^2}-\Xi\frac{\mcl H^2}{\phi'^2}\lsb \lb
1+\frac{\mcl H'}{\mcl H^2}-\frac{\phi'^2}{2\mcl H^2}-\frac{2\phi''}{\mcl
H\phi'}-\frac{4\Ka}{\mcl H^2}\rb+\frac{k^2}{2\mcl H^2}\lb
P_++P_-\rb\rsb-\frac{\mcl H^2}{\phi'^2}\frac{k^2}{2\mcl H^2}\lb
P_++P_-\rb,\nn\\ \mcl C_{31}&=&8+2\frac{\mcl H^2}{\phi'^2}\lsb
2\frac{\phi^{(3)}}{\mcl H^2\phi'}+5\lb \frac{\mcl H'}{\mcl
H^2}-1\rb-2\frac{\phi''}{\mcl H\phi'}\lb 1+\frac{\phi''}{\mcl
H\phi'}\rb\rsb,\nn\\ \mcl C_{32}&=&\frac{\mcl
H^2}{\phi'^2}\frac{k^2}{\mcl H^2}\lsb\frac{2}{3}-\frac{4}{3}P_++2\Xi\lb
\frac{\mcl H'}{\mcl H^2}-1\rb-\frac{\Xi}{3}\rsb.\nn \eea
}
\end{document}